\documentclass[useAMS,usenatbib]{mn2e}
\usepackage{epsfig,float,natbib,amsmath,longtable,supertabular}

\title[Monthly Notices: \LaTeXe\ guide for authors]
  {Monthly Notices of the Royal Astronomical
  Society: \\ \LaTeXe\ style guide for authors}

  \title[Crossing into the substellar regime in Praesepe]
  {Crossing into the substellar regime in Praesepe}
\author[R. J. Chappelle et al.]
  {R. J. Chappelle$^1$, D.J. Pinfield$^{1,2}$, I.A. Steele$^1$, P. D. Dobbie$^3$, A. Magazz\`{u}$^4$ \\
  $^1$Astrophysics Research Institute, Liverpool John Moores University, Twelve Quays House, Egerton Wharf, Birkenhead, CH41 1LD, UK\\
  $^2$Department of Physics, Astronomy and Mathematics, University of Hertfordshire, College Lane, Hatfield AL10 9AB, UK\\
  $^3$XROA Group, Department of Physics and Astronomy, University of Leicester, University Road, Leicester LE2 7RH, UK\\
  $^4$Fundaci\'{o}n Galileo Galilei-INAF, Apartado 565, E-38700 Santa Cruz de La Palma, Spain.;\\
      Osservatorio Astrofisico di Catania, Via S. Sofia 78, I-95123 Catania, Italy}
\pagerange{\pageref{firstpage}--\pageref{lastpage}} \pubyear{2004}

\def\LaTeX{L\kern-.36em\raise.3ex\hbox{a}\kern-.15em
    T\kern-.1667em\lower.7ex\hbox{E}\kern-.125emX}

\begin{document}

\label{firstpage}

\maketitle

\begin{abstract}
We present the results of a deep optical 2.6 square degree survey with near infrared follow-up measurements of the intermediate-aged Praesepe open cluster. The survey is complete to \textit{I$_c$}=21.3, \textit{Z}=20.5, corresponding to M$\sim$ 0.06 M$_\odot$ assuming a cluster age of 0.5 Gyrs. Using 3-5 pass-bands to constrain cluster membership, we identify 32 new low mass cluster members, at least 4 of which are likely to be substellar. We use the low mass census to trace the region where the sequence moves away from the NEXTGEN towards the Dusty regime at T$_{eff}$ = 2200K. In doing so, we identify four unresolved binaries, yielding a substellar binary fraction of $\sim$ 30 percent. The binary fractions appear to decrease below 0.1 M$_\odot$, in contrast to the rising fractions found in the Pleiades. We also identify a paucity of late M dwarfs, thought to be due to a steepening in the mass-luminosity relation at these spectral types, and compare the properties of this gap in the sequence to those observed in younger clusters. We note an overdensity of faint sources in the region of the so-called subcluster (possibly an older smaller cluster within Praesepe), and subsequently derive the luminosity and mass functions for the main Praesepe cluster, revealing a turn-over near the substellar boundary. We conclude by presenting astrometric measurements for low mass Praesepe candidates from the literature, and rule out as a likely foreground dwarf RPr1, hitherto thought to be a substellar member.    
\end{abstract}

\section{Introduction}
There has been much work done surveying young open clusters for very low mass stars (VLMS) and brown dwarfs (BDs) (e.g. \citealt{bouv_98}, \citealt{dob_hyades}). The focus has generally been to identify BDs and measure cluster mass functions (MFs) in a variety of environments to establish if the initial mass function (IMF) is universal or not -- the answer to which has important implications for our understanding of star and BD formation. The study of older clusters is also important if we are to understand how the dynamical evolution of the cluster affects the shape of their MF -- over time VLMS and BDs are expected to be preferentially ejected (eg. \citealt{fuente_marcos}). For the Hyades, \citet{dob_hyades} searched for low-mass (0.1--0.06M$_{\odot}$) members in a 10.5 square degree survey, but found only one (previously known) stellar member. They estimate that 4--5 Hyads should have been found in this mass range if the Hyades and Pleiades MFs are identical, and conclude that dynamical evaporation of low-mass Hyads is the most likely explanation for the deficit. However, studies of other clusters with similar ages are clearly needed if we are to properly address this question.
\par
Late cluster members are also ideal for testing ultra-cool atmosphere models of late M and L spectral types (T$_{eff}$ $\leq$ 2500K \citealt{legg_01}). At such low T$_{eff}$, atmospheric dust grains condense out of the gas phase, and strongly affect both colour and spectral properties. However, dust grain properties depend not only on temperature but on gravity ($g$)and metallicity ([M/H]) which cannot be measured with confidence for late M and L field dwarfs. But with well constrained age (and hence radii and masses inferred from evolutionary models) late M and L cluster members' $g$, T$_{eff}$ and [M/H] can be known. Indeed, spectroscopic studies of such objects with a range of different (but known) ages would allow us to empirically isolate the T$_{eff}$, $g$, and [M/H] dependence of both narrow and broadband spectral features.
\par
The Praesepe cluster has an age of (0.9$\pm$0.5) Gyrs, lies at a distance of $\sim$ 170 pc, has a near solar metallicity \citep{hshja} and zero reddening \citep{crawford69}. Large scale proper motion studies of Praesepe include \citet{hshjb} (hereafter HSHJ) covering 19 sq. degs down to 0.1M$_\odot$, and \citet{adams02} covering the whole cluster to similar depth. A number of smaller but deeper Praesepe surveys have also been carried out. \citet{pin_prae} (P97 hereafter) covered $\sim$ 1 sq. deg down to $I_c$ = 21. \citet{pin_t} (P00 hereafter) covered $\sim$ 5 sq. degs down to \textit{I}$\sim$19.5. \citet{mag} covered 800 sq. arcminutes down to \textit{I}=21.2, and discovered one candidate member of M9 spectral type (Roque Praesepe1; RPr1 hereafter). If confirmed as a Praesepe member, its estimated mass would be M=0.063--0.084$M_{\odot}$ making it a possible BD. NIR characterisation of the faint candidates has been presented by \citet{hodge} and \citet{pin_2003} (P03 hereafter).
\par
In this paper, we present the results of a deep 2.6 square degree optical survey of the cluster, and use near-infrared follow-up measurements to refine photometric membership status. We then combine our candidates with others from the literature, and identify likely unresolved binaries. We also recover the ``M dwarf gap'' in the cluster -- a dearth of M7-8 dwarfs previously noted in the Pleiades and other clusters (\citealt{dobbie}; P03). We derive the luminosity function, use state-of-the-art evolutionary models to estimate candidate masses and determine the cluster mass function. We also derive proper motions for several previously identified cluster candidate members. Finally, we discuss planned future work.
\section{WFC \textit{IZ} SURVEY}
\subsection{Observations}
Deep \textit{I} and \textit{Z} band images of Praesepe were obtained with the INT/WFC during the nights 2001/12/24-28. The WFC instrument consists of 4 thinned 2048 x 4196 pixel CCDs (0.333 arcseconds/pixel), and operates at the prime focus of the INT, covering a 0.29 sq. deg field of view.
\par
\begin{figure}
\includegraphics[width=84mm]{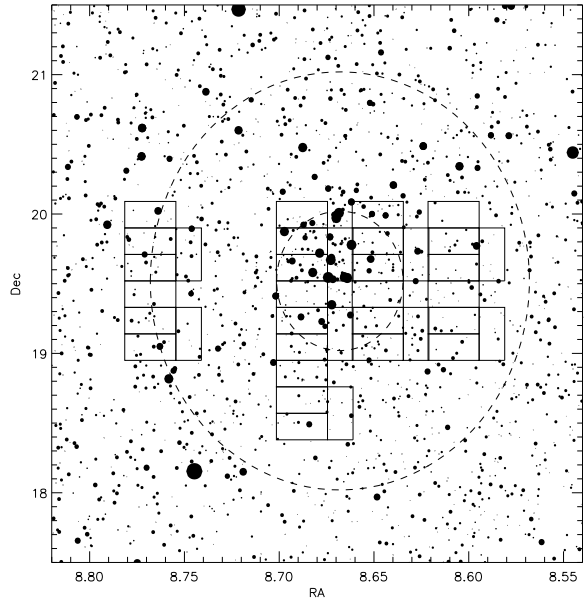}
\caption{INT survey sky coverage. The 9 WFC fields are shown. 2 circles of radii 0.5 and 1.5 deg centred on the cluster centre are also shown. Point sources from the USNO catalogue are shown as filled circles, with size indicating apparent magnitude.} \label{fig:sky_cov}
\end{figure}

The conditions for $\sim$ 35 percent of the time were photometric with modest humidity and seeing of $\sim$ 1 arcsec. During the run, we observed 9 pointings through the \textit{I}$_{rgo}$ and \textit{Z}$_{rgo}$ filters with exposure times of 20 minutes per band. In total, 2.6 square degrees of the cluster were surveyed. Our coverage is shown in Figure \ref{fig:sky_cov}.

\subsection{\textit{IZ} reduction} 
Data reduction was performed using standard IRAF\footnote{IRAF is distributed by the National Optical Astronomy Observatories, which are operated by  the Association of of Universities for Research in Astronomy, Inc., under cooperative agreement with the National Science Foundation.} routines. The images were bias subtracted, and non-linearity accounted for prior to flat-fielding. Fringe maps, which were constructed in both bands by median filtering images acquired throughout the entire run, were used to remove the effects of the interference between night sky lines in the CCD substrate.
\par
 Sources were identified in the \textit{I}-band images using IRAF'S DAOFIND at a detection level of  $3 \sigma$, and matched to \textit{Z}- counterparts using a 4 pixels search radius of the \textit{I}-band source position (taking into account any systematic shifts between the \textit{I}- and \textit{Z}- images). Aperture photometry was measured for each source using the PHOT routine with a 7 pixel aperture size and a sky value computed from within a circular annulus.
\par
To calibrate the photometric data, standard stars drawn from \citet{land} were observed, and zero points and extinction coefficients determined. In order to calibrate the 7 images obtained in non-photometric conditions, we re-observed these fields with shorter (45s) integrations in the \textit{I-} and \textit{Z-} bands on the night of 23rd February 2002 during photometric conditions. Landolt standard observations were used to calibrate the magnitudes of $\sim$ 50 of the brighter stars in each of these shorter integrations, and of these, the $\sim$ 40 stars that were not saturated in our longer integrations were used as secondary standards to determine zero points for these longer exposure images.  
\par

\subsection{WFC Results}

\begin{figure}
\includegraphics[width=84mm]{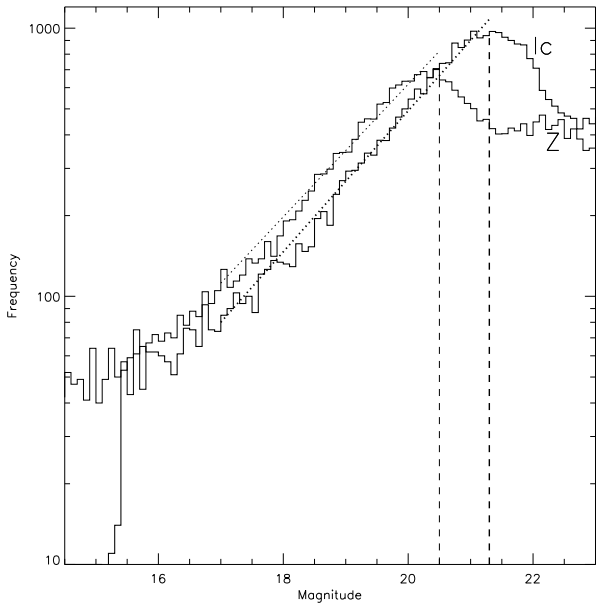}
\caption{\textit{I} and \textit{Z} magnitude histograms of source counts in the Praesepe WFC survey. \textit{I-} (bold) and \textit{Z-} band data are shown by solid lines. Dashed lines indicate the 90 percent completeness limits.} \label{fig:iz_hist}
\end{figure}

Figure \ref{fig:iz_hist} plots the frequency of the total number of sources detected per 0.1 magnitude interval in the \textit{I}- and \textit{Z}-bands. The dotted lines are straight line fits to the \textit{I-} and \textit{Z-} band histograms in the magnitude ranges \textit{I}=17-20.5 and \textit{Z}=17-19.5. Survey completeness limits for the \textit{I} and \textit{Z} bands were taken to be the magnitudes at which the binned star counts drop below 90 percent of the expected frequency, and are shown as vertical dashed lines. In this way, the survey was estimated to be 90 percent complete to \textit{I} =21.3, \textit{Z}=20.5.
\par
\begin{figure}
\includegraphics[width=84mm]{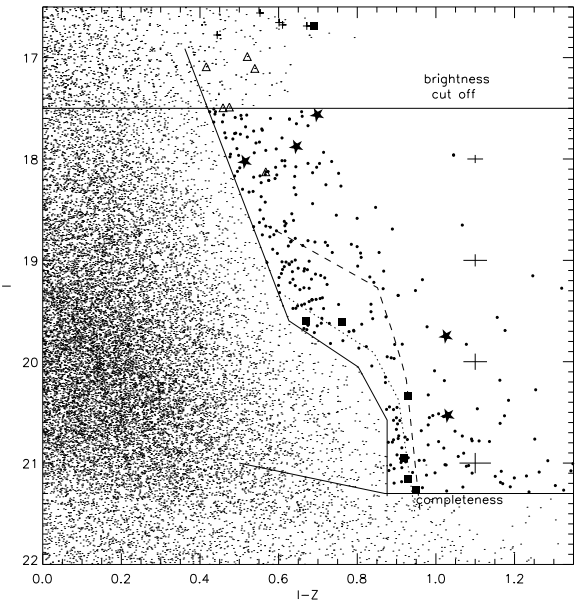}
\caption{WFC \textit{IZ} colour magnitude diagram. The solid line indicates the redness cut-off for Praesepe members. The dashed and dotted lines show the theoretical isochrones for the 0.5 Gyr DUSTY and NEXTGEN models respectively. Plus signs and triangles are cluster LMS from HSHJ and P03 surveys respectively. Proper motion members identified in this work are shown as filled stars. The expected positions of field dwarfs (from \citealt{cossburn}) are shown as squares.} \label{fig:iz_cm}
\end{figure}

Our  WFC survey resulted in a catalogue of $\sim$ 15000 objects. The resultant \textit{I},\textit{I}-\textit{Z} colour-magnitude diagram (CMD) is shown in Figure \ref{fig:iz_cm}. In order to separate candidate cluster members from background sources, it was necessary to locate the cluster sequence. To do this, we used 0.5 Gyr DUSTY \citep{chab2000} and NEXTGEN \citep{bar98}  model isochrones (dashed and dotted lines respectively), transformed onto our CMD. Predicted \textit{I}-\textit{Z} colours were calculated using the relations given in \citet{dobb02}, where we assume a distance modulus of (m-M)$_0$ = 6.16 \citep{pinn_hip} and zero extinction. To further highlight the location of the cluster sequence, we also identified HSHJ proper motion members and optical/NIR candidates from P03 that were recovered in our survey. We choose not to include objects that fall within 1 degree of the centre of a suspected sub-cluster, discussed in $\S$\ref{sec:mf}. These are highlighted in the figure as plus signs and triangles respectively. Further, we overplot candidates from P03 that are shown to be astrometric cluster members in section $\S$\ref{sec:astrometry}, and are shown as filled stars.
\par
Also plotted in Figure \ref{fig:iz_cm} are the expected magnitudes of several M field dwarfs as they would appear at the cluster distance. Praesepe members are sufficiently old to have finished their contraction stage, implying that cluster UCDs will have the same luminosity as field stars of the same T$_{eff}$. The magnitudes for these are taken from \citet{cossburn}, and are shown as squares in the CMD.  
\par
Using the photometric and astrometric members, combined with the offset field star positions and the synthesized model tracks as a guide on the CMD, we have defined a cut-off line to separate background sources from potential cluster members. By taking into account photometric uncertainties (indicated on the right hand side of the figure), we have positioned our cut-off line such that it is at least 1-sigma (in I-Z uncertainty) blueward of these photometric guides. Note that for I$>$19, the offset field stars are the closest objects to the cut-off line, and the proper motion cluster members are rather redder than the cut-off. This is
as expected, since Praesepe low-mass members (Teff$<$2500K) will not be fully contracted and we expect lower Teffs as a function of magnitude (e.g. ~100K lower for Mi~14 at 0.5Gyr age), and thus slightly larger I-Z colours (e.g. ~0.1 larger; \citealt{dobbie_iz}). The use of the offset field stars thus provides a conservative way to define our cluster region on the CMD. We are thus confident that our selection will not miss any genuine cluster members. We further constrained the selection of candidates using a bright limit (brighter than which 2MASS is sensitive to cluster members which will have been previously identified by \citet{adams02} and HSHJ), and a faint completeness limit (both indicated on the CMD). All stellar-like objects redward of the cut-off line, and between the bright and faint limits were selected as \textit{IZ} candidates, and are shown as filled circles in the figure. We identified 320 cluster candidates in this way from $I_c$ = 17.5-21.3. Visual inspection of each revealed 95 to be spurious (mostly diffraction spikes and cosmic ray hits), leaving 225 genuine candidate members.

\section{NIR FOLLOW UP}
Near-infrared follow-up measurements of the 225 WFC candidate members were made using the Fast Track Imager (UFTI) instrument on the United Kingdom Infrared Telescope (UKIRT). UFTI was windowed (512x512 pixels) for faster readout providing a 46.5 arcsecond field of view, with each integration consisting of a 5-point dither pattern. \textit{K-}band photometry of the 120 brightest candidates was measured during flexibly scheduled queue observing in photometric conditions from the 19th to 24th of February 2003. \textit{K}-band photometry of the remaining 105 fainter candidates, as well as longer \textit{J}-, \textit{H}- and \textit{K}-band integrations were obtained by us in photometric conditions during our UKIRT observing run on 7-10 March 2003.
\par
Our observing strategy was to initially measure K-band photometry (S/N$\sim$20) for all candidates, with exposure times ranging between 6s to 60s depending upon source brightness. At this stage we considered a target as a potential cluster member if it lay near to the NEXTGEN isochrone for \textit{K}$\le$15.5, had \textit{I}-\textit{K}$>$3.3 for \textit{K}=15.5--16.5, or had \textit{I}-\textit{K}$>$3.8 for \textit{K}$>$16.5. We then obtained longer integration \textit{J}, \textit{H} and \textit{K} measurements of 12 of the faintest potential cluster members, where we used exposure times of 24s-120s, aiming for a S/N of $\sim$ 50. The optical and NIR photometry of all WFC sources flagged as potential candidate members is presented in Table \ref{tab:photom}.
\par
The images were de-biased, dark subtracted, flat fielded and finally combined into mosaics using the UKIRT software pipeline ORACDR \citep{bridger}. The STARLINK package GAIA was then used to measure the photometry, using apertures matched to the seeing (0.6--1 arcseconds). Individual aperture corrections were calculated using UKIRT faint standards measured either side of the candidate, and instrumental magnitudes were then transformed onto the MKO system using airmass curves derived from these standards.
\section{ADDITIONAL PRAESEPE CANDIDATES}
As well as the new WFC candidates, we also consider Praesepe candidate members from several other surveys. Proper motion candidates from HSHJ define the bright main sequence down to \textit{I}$_c$ $\sim$ 17.5. We have transformed the HSHJ photographic \textit{I}$_N$ magnitudes onto the Cousins system using \citet{bessell}. Sources from P97 and P00 (which we will refer to as \textit{Riz} and \textit{Iz} candidates respectively) were also considered in the present analysis. NIR photometry of the \textit{Riz} and \textit{Iz} candidates was taken from P03, and is on the MKO system. We also considered the RPr1 candidate from \citet{mag} (the M98 survey hereafter). The NIR photometry of this object has been transformed onto the MKO system using the transformations of \citet{hawarden}.
\par
In order to obtain NIR measurements of as many of the HSHJ stars as possible, we cross matched their positions with the 2MASS All Sky catalog. 2MASS photometry typically provides a SNR$\sim$10 for \textit{J} = 15.8, \textit{H} = 15.1, and \textit{Ks} = 14.3. We identified NIR counterparts within a search radius of 3 arcseconds from each source. Of the 459 HSHJ candidates that have astrometric membership probabilities $\geq$ 70 percent (from HSHJ), 447 had a corresponding 2MASS counterpart. 3 of these were flagged by 2MASS as having potentially contaminated photometry. These, and the 12 unmatched HSHJ candidates, are noted in Table \ref{tab:photom_hshj}. We transformed the 2MASS photometry onto the MKO system using equations in \citet{carp}. 
\section{PHOTOMETRIC MEMBERSHIP AND BINARITY} 
\subsection{Membership from the \textit{IK} CMD}\label{phot_res} 
\begin{figure*}
\includegraphics[width=124mm]{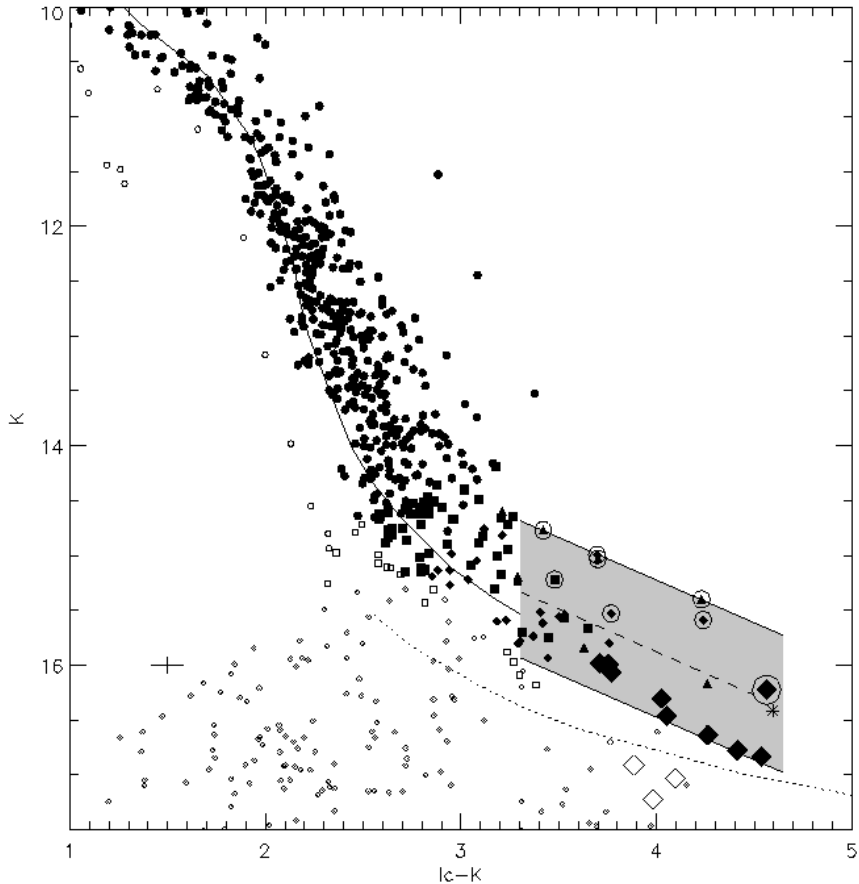}
\caption{The Praesepe \textit{K},\textit{I}$_c$ CMD. Filled symbols represent cluster candidates (binary candidates are circled), and empty symbols are photometrically rejected candidates. Circles are HSHJ candidates, squares are \textit{Iz} candidates and triangles are \textit{Riz} candidates. For the current survey, diamonds are \textit{K} band only WFC candidates, and large diamonds are \textit{JHK} WFC candidates. The solid and dotted isochrones are 0.5 Gyr NEXTGEN and Dusty models respectively. RPr1 is shown as an asterisk. The greyed area and dashed line within defines the \textit{cluster sequence region}, and is discussed in the text.} \label{fig:cm_k}
\end{figure*}

Figure \ref{fig:cm_k} shows the \textit{K},\textit{I-K} CMD for Praesepe. \textit{Iz} and \textit{Riz} candidates from P97 and P03 are shown as triangles and squares respectively. The new WFC candidates are shown as diamonds. Large diamonds represent WFC candidates that were observed using longer \textit{K}, \textit{J} and \textit{H} integrations. All candidates with photometry which we deem inconsistent with cluster membership (see below) are displayed as empty symbols. RPr1 is shown as an asterisk. 
\par
Over-plotted on the CMD are 0.5 Gyr NEXTGEN (solid line) and Dusty (dotted line) model isochrones from the Lyon Group (\citealt{bar98}, \citealt{chab2000}), where we have assumed a distance modulus of (m-M)$_o$ = 6.16 and zero reddening.
\par
The large number of HSHJ and \textit{Iz} sources define the sequence well down to \textit{K}$\sim$15.5, and it is clear from the figure that the NEXTGEN isochrone fits the cluster sequence fairly well down to this magnitude (out to \textit{I}-\textit{K}$\sim$3.3; as was noted by P03). We therefore adjudged the membership status of all candidates with \textit{I}-\textit{K} $\leq$ 3.3 based on their proximity to the NEXTGEN isochrone. Candidates which appeared to be too blue, despite making allowance for the cluster depth (+/-0.15 mags; \citealt{holland}) and the photometric uncertainties, were rejected.
\par
All 444 HSHJ objects with 2MASS photometry have \textit{I}-\textit{K} $\leq$ 3.3. Using the NEXTGEN isochrone as a test for cluster membership, we adjudge 18 of these to be non-members. Furthermore, three HSHJ candidates appear significantly redder than the isochrone, with photometry indicative of foreground field dwarfs. However, we caution that these sources are all found on the same Schmidt plate in HSHJ suggesting that their redness could be due to calibration systematic errors, so we choose not to reject them as candidates. In Table \ref{tab:photom_hshj} we list these 3 HSHJ candidates, the photometric non-members and the HSHJ candidates for which no 2MASS counterpart was found or a 2MASS contamination flag was logged. All other HSHJ candidates have 2MASS photometry consistent with cluster membership. 
\par
For \textit{I}-\textit{K} $\geq$ 3.3 (T$_{eff}$ $\leq$ 2500K) the cluster sequence is not obviously defined by the candidates themselves. Also it is not immediately apparent if either the NEXTGEN or DUSTY isochrones are appropriate. This is the T$_{eff}$ range when atmospheric dust grains are expected to condense out of the gas phase \citep{jones_97}, so we might expect the NEXTGEN isochrone to be inadequate. However, the DUSTY isochrone is clearly too blue when one moves fainter than \textit{K}=15.5, and this isochrone is thus not a viable alternative to guide candidate member selection. Consequently, we chose not to rely on the models to define our Praesepe sequence in this range.
\par
Instead, we defined a \textit{cluster sequence region} in which to identify candidates as likely members. In order to do this we began by defining the top of the equal mass unresolved binary sequence in the \textit{K},\textit{I-K} CMD, which we assume consists of the brightest candidates in this colour range -- these sources then define the bright limit of our cluster sequence region. The single star sequence will be 0.75 magnitudes fainter than the equal mass binary sequence. We allow for the broadening of both the single and binary sequences by cluster depth effects and assume photometric uncertianties of +/-0.2 magnitudes are applicable to both single and binary members. Therefore, as illustrated by the shaded zone in Figure \ref{fig:cm_k}, our bright and faint selection criteria are separated by 1.45 magnitudes.  We adjudge candidates with \textit{I}-\textit{K} $>$ 3.3 lying outwith this region as non-members.
\par
To summarise, of the 225 WFC candidates with \textit{K}-band measurements, 90 look like non members, and 36 have \textit{I}-\textit{K} colours consistent with cluster membership (84 percent contamination amongst the candidates from the I, I-Z CMD). Of the 12 WFC candidates for which we measured long integration \textit{J}, \textit{H}, and \textit{K} magnitudes, 3 are now flagged as non members, because although red (and initially considered interesting) they lie below our cluster sequence region in the CMD. Our \textit{IK} membership criteria are given in Table \ref{tab:photom}, and all \textit{IK} members (including those from the other surveys considered) are listed in Table \ref{tab:photom_red}.
\subsection{Identifying Unresolved Binarity}
\label{subsec:unres_bin}
Possible unresolved cluster binaries were identified in the \textit{I}, \textit{I}-\textit{K} CMD by dividing our cluster sequence region in two, using a division that is mid-way between the region's bright and faint limits. This division is shown in Figure 4 as a dashed line. Candidate cluster members above this line will lie closer to the equal mass binary sequence than to the single star sequence, and have been duly flagged as binary candidates (these sources are circled in Figure 4). Unresolved binarity amongst the \textit{Riz} and \textit{Iz} candidates (as well as the HSHJ sample) has been previously addressed by P03. We have nothing significant to add to their analysis of candidate members with \textit{I}-\textit{K}$<$3.3. However, our current sample of redder candidates is larger and goes fainter than P03. Therefore we have re-assessed the unresolved binarity of any \textit{Riz} and \textit{Iz} candidates with \textit{I}-\textit{K}$\ge$3.3 (as well as the WFC candidates) using our current approach. Unresolved probable binaries are flagged in the last column of Table A3.
\subsection{The \textit{JK} CMD}
\begin{figure}
\includegraphics[width=84mm]{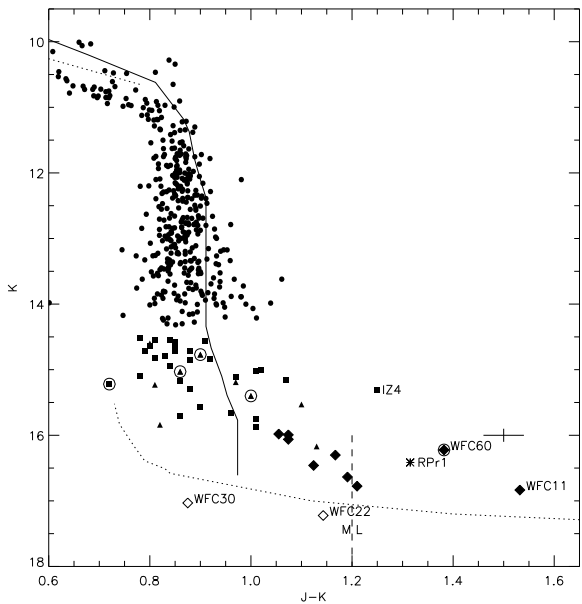} 
\caption{The Praesepe \textit{K}, \textit{J}-\textit{K} CMD. Symbols are the same as in figure \ref{fig:cm_k}. The M-L spectral boundary, estimated from \citet{legg_98} and \citet{knapp}, is indicated by a dashed line at \textit{J-K}=1.2. WFC candidates discussed in the text are labelled.} \label{fig:cm_jk}
\end{figure}
The \textit{K},\textit{J}-\textit{K} CMD is shown in Figure \ref{fig:cm_jk}. All faint candidates with accurate \textit{J}-band measurements are plotted. We only include HSHJ sources with \textit{K-} uncertainties $<$0.06 magnitudes, and we do not show {\it Iz} candidates with photometry from 2MASS due to the large photometric uncertainties. NEXTGEN and DUSTY model isochrones for 0.5Gyr age are shown as solid and dotted line respectively. It can be seen that the NEXTGEN isochrone agrees well with the cluster members to \textit{K} $\approx$ 15.5 (as with the IK CMD). However, for \textit{K}$>$15.5 the candidates move significantly red-ward of the NEXTGEN isochrone. In the main, these fainter candidates lie slightly above the DUSTY isochrone. However, the NIR colour trend shows the rapid increase in \textit{J}-\textit{K} that the DUSTY models predict, although it occurs at a slightly brighter magnitude.
\par
Three candidates with \textit{J}-band photometry were assigned as non-members in \S 5 (two of which appear on the plot as open symbols). WFC30 and WFC22 lie slightly below the dusty isochrone but have \textit{J}-\textit{K} colours that are comparable with the majority of faint candidate members. This is consistent with these two sources being background stars. WFC57 has a very red colour (\textit{J-K}=1.91) and lies redward of the plotting area. Although this is suggestive of a late L dwarf, the \textit{I-K} colour of 3.88 is not red enough to be a late L dwarf (cf. \textit{I-K}=4--5). Consequently we believe this source is a red galaxy. \textit{Iz}4 is also highlighted in Figure \ref{fig:cm_jk}, and appears to look like a bright foreground field dwarf, but the possibility of it being an unresolved binary cannot be ruled out. Thus at this stage we do not discard it as a contaminant.

\par
The approximate M/L transition colour is indicated at the bottom of the plot (estimated using colours from \citealt{legg_98} and \citealt{knapp}), and shows that most of our faint Praesepe candidate members should be very late M dwarfs. At least two however (WFC 11 and WFC60), have colours consistent with being early-mid L spectral types. These are labelled in Figure \ref{fig:cm_jk}. WFC60 was flagged as an unresolved binary in \S 5.2, and in the \textit{JK} CMD it is at least 0.5 magnitudes brighter than candidate members with similar \textit{J-K} colours. This candidate could be an unresolved binary L dwarf in the cluster.
\subsection{Contamination}
\label{sec:contamination}
In this section we estimate the expected level of field contamination at the faint end of the Praesepe sequence. We do this for three \textit{K-} band magnitude ranges: 15.5-16.0, 16.0-16.5 and 16.5-17.0. In the cluster member region of the \textit{IK} CMD, it can be seen that these ranges correspond to \textit{I-K} colours of $\sim$ 3.5, 4.0 and 4.4 respectively. 
\par
These colours relate to spectral types $\sim$ M6, M8 and L0 respectively. The absolute magnitudes of field stars with these spectral types (from \citealt{dahn} and \citealt{knapp}) means those at 160-200 pc will overlap the cluster sequence in a CMD. 
\par
The sky area covered for each magnitude range varies, since the different surveys considered have different photometric depth. All of the surveys considered (excepting the HSHJ survey) contribute to the brightest magnitude range. However, the \textit{Iz} survey does not contribute to the two fainter ranges. All except the HSHJ and \textit{Iz} surveys contribute to the middle magnitude range, and only the WFC and M98 surveys contribute to the faintest magnitude range. After accounting for overlap between certain survey areas, we find that the brightest magnitude range covers a volume of $\sim$ 1.2 x 10$^3$ pc$^3$, and the two fainter ranges each cover a volume of 1.0 x 10$^3$ pc$^3$.
\par
Luminosity functions predict late M and early L field dwarfs densities of $\Phi$ = $\sim$ $2.5$ x 10$^{-3}$ and $1.9$ x $10^{-3}$ stars $pc ^{-3}$ (\citet{kirk_94} and \citet{cruz} respectively). We hence expect a field contamination of $\sim$ 3 late M dwarfs and $\sim$ 2 early L dwarfs amongst the candidates.
\par
While clearly the levels of contamination amongst our candidate members with the late M spectral types is fairly low ($\sim$20\%), it is entirely possible that both of our candidate L type members could be field stars. A more rigorous assement of their membership status can be performed by measuring their proper motions.

\subsection{\textit{J}-\textit{H},\textit{H}-\textit{K} 2 colour diagram}
\label{sec:cc}
\begin{figure*}
\includegraphics[width=124mm]{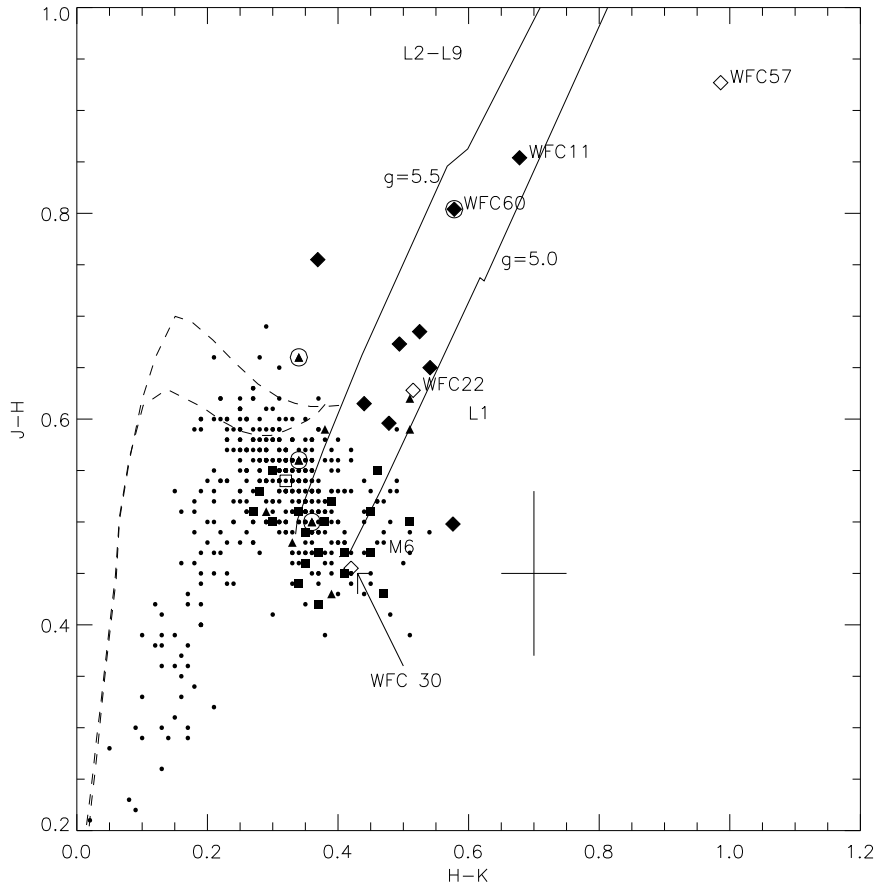} 
\caption{The Praesepe \textit{J}-\textit{H},\textit{H}-\textit{K} 2-colour diagram. Symbols are the same as in figure \ref{fig:cm_k}. Spectral type estimates from \citet{leg02} are overplotted.} \label{fig:cc_plot}
\end{figure*}
Figure \ref{fig:cc_plot} shows our candidates in the \textit{J}-\textit{H},\textit{H}-\textit{K} 2-colour diagram. The symbols are the same as in the CMDs. NEXTGEN (dashed lines) and DUSTY (solid lines) model colours are over-plotted, with the surface gravity values shown. The bluer synthetic \textit{H}-\textit{K} colours for \textit{H}-\textit{K}$<$0.2 are thought to be due to inaccuracies with water vapour modelling in the H band. Approximate late M and L spectral class locations are also indicated (from \citet{legg_98} and \citealt{knapp}).
\par
Visual inspection of the diagram shows that, as expected, the 2 likely background stars (WFC 22 and 30) have colours consistent with late M dwarfs. WFC57, probably a galaxy, has NIR colours similar to a late L dwarf. The possible unresolved L dwarf binary (WFC60) and the even redder single candidate member (WFC11) have NIR colours consistent with being early-mid L dwarfs. Similarly, \textit{Riz}117 appears too blue for an early L dwarf, and is therefore considered as a probable non-member. It is shown in Figure \ref{fig:cc_plot} as an open square.
\section{THE PRAESEPE SEQUENCE}
\subsection{M dwarf gap}
\label{sec:gap}
In Figure \ref{fig:cm_k_emp} we show a zoomed in portion of the Praesepe \textit{IK} CMD with non-members omitted. The four binary tracks (shown with lines joined by plus signs) will be discussed in \S \ref{sec:bin_mf}.
\par
It can be seen that there is a dearth of cluster candidates from \textit{K}=15.3-15.5. \citet{dobbie} and P03 presented photometric evidence for a similar paucity of M7-M8 spectral types in other clusters, as well as in star forming regions and the field. This paucity has been referred to as the ``M dwarf gap'' (P03). The fact that the M dwarf gap has been observed in both young associations such as the Pleiades and $\theta$ Orionis as well as amongst older populations (M4 and the field), suggests that its origins are due to a sharp local increase in the slope of the magnitude-mass relation for late-M spectral types ($T_{eff} \approx 2700 K$). At this T$_{eff}$, models predict that dust grains begin to condense in the outer atmospheric layers (e.g. \citealt{tsuji}), and as \citet{dobbie} suggests, grain opacities could be responsible for the gap feature. The Dusty models do not consider grains with sizes greater than $\sim$ 0.24 $\mu m$, and consequently predict negligible grain opacities in the NIR, and no gap. Grain formation models however, suggest that grains could have sizes up to tens of microns \citep{cooper}, which could result in a significant opacity increase due to Mie scattering.
\par
We have measured the extent of Praesepe's ``M dwarf gap'' in the \textit{I-},\textit{J-},\textit{H-} and \textit{K-} bands. For each band, the gap size was calculated as the magnitude difference between the faintest candidate above the gap, and the brightest candidate below, ensuring that only the candidates with full \textit{IJHK} photometry were used. The results are tabulated (with the corresponding values for the Pleiades from P03) in Table \ref{tab:mdg}. For all passbands, quoted errors were derived by combining the photometric uncertainties of the two candidates that were used to measure the gap sizes. 
\par
It can be seen that the gap is smaller for Praesepe than for the Pleiades in all the bands considered, although there is a small overlap if the uncertainties are taken into account. This is consistent with the interpretation just mentioned, since (as \citet{dobbie} explains) one expects a shallower luminosity-T$_{eff}$ relation for older populations (such as Praesepe), because the radius-mass relation will be steeper for younger populations (like the Pleiades), where objects are still in the relatively early stages of contraction. The relative extent of the Praesepe M dwarf gap in the four photometric bands considered is comparable for both clusters. We will, however, discuss more subtle colour changes across the gap in the next section.
\begin{table}
\caption{Sizes of the ``M dwarf gap'' in the \textit{I}-,\textit{J-},\textit{H-} and \textit{K-} bands for the Pleiades and Praesepe}
\label{tab:mdg}
\begin{tabular}{|c|c|c|c|c|c|}
\hline
Cluster  & \textit{I}  & \textit{J}  & \textit{H}  & \textit{K} \\
\hline
\hline
Pleiades & 0.5$\pm$0.05  & 0.3$\pm$0.06   & 0.3$\pm$0.04   & 0.3$\pm$0.05  \\
\hline
Praesepe & 0.30$\pm$0.03   & 0.25$\pm$0.05   & 0.24$\pm$0.05   & 0.23$\pm$0.03   \\
\hline
\end{tabular}
\end{table}
\subsection{The faint single star sequence}
\begin{figure*}
\includegraphics[width=100mm]{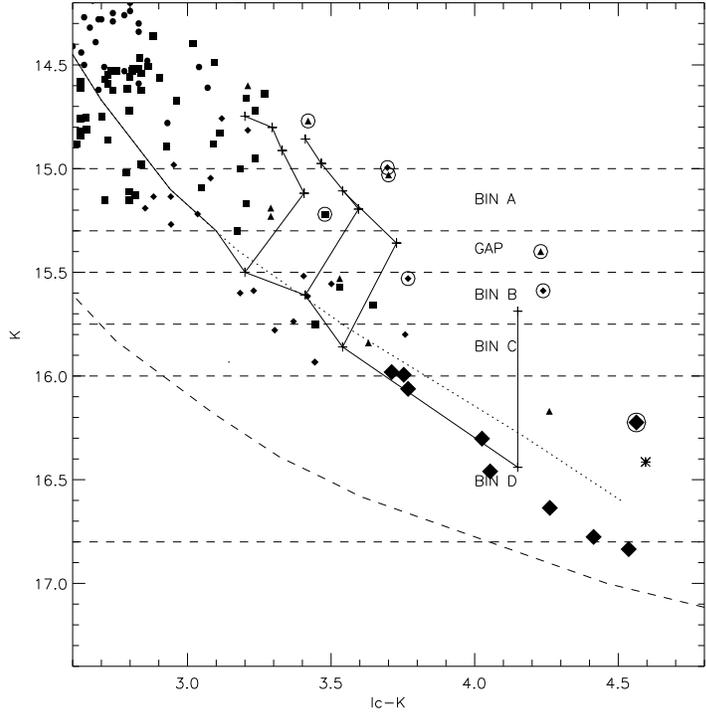}
\caption{\textit{K}, \textit{I}-\textit{K} CMD. Symbols are the same as for figure \ref{fig:cm_k}. Solid line shows the empirical sequence that tracks the move away from the NEXTGEN isochrone and onto the dusty sequence. 0.5 Gyr NEXTGEN and Dusty models are shown as dotted and dashed lines respectively. 4 binary tracks are also shown.} \label{fig:cm_k_emp} 
\end{figure*}
In order to study how the bulk properties of the Praesepe candidates change for $K>$15, we split the Praesepe candidate members into the following \textit{K}-band magnitude bins: bin A (15.0 -- 15.3) represents the candidates slightly brighter than the M dwarf gap (discussed in the last section). Bins B and C (15.5--15.75 and 15.75--16.00 respectively), are below the gap. Bin D covers all the fainter candidates (except WFC 11) down to \textit{K} =16.8. These 4 bins are indicated by dashed lines in Figure \ref{fig:cm_k_emp}.
\par
In order to determine an ``observational photometric sequence'', we used the NEXTGEN model isochrone above the gap region in the CMD (since this model agrees well with the data in this range). To bridge the M dwarf gap we have used our measured gap sizes (see Table 1) to provide colour and magnitude changes from K=15.3 -- 15.5. Below the gap (bins B--D), we have averaged the colours and magnitudes of our single star candidate members in each bin. Our ``observed cluster sequence'' is shown as a solid line in Figure \ref{fig:cm_k_emp}. 
\par
When comparing the observed sequence to the model isochrones, it is apparent that for \textit{I}-\textit{K}=3.3--4 the NEXTGEN model is still quite reasonable. However, for \textit{I}-\textit{K}$>$4 the candidates appear to drop slightly below the NEXTGEN isochrone, moving closer to the DUSTY isochrone. P03 and \citet{jame} pointed out a similar transition from the NEXTGEN to DUSTY isochrones for the Pleiades cluster sequence from \textit{I}-\textit{K}=4.1--4.5.
\begin{figure*}
\includegraphics[width=100mm]{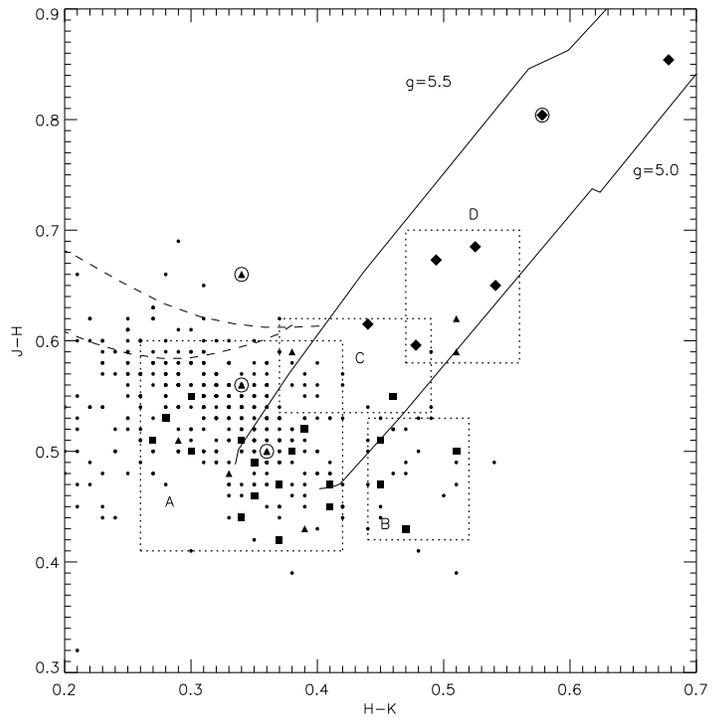}
\caption{\textit{J}-\textit{H},\textit{H}-\textit{K} 2-colour diagram. Symbols are the same as for figure \ref{fig:cm_k}. Dotted lines define the perimeters of 4 \textit{K} magnitude bins, discussed in the text. } \label{fig:cc_emp} 
\end{figure*}
\par
We trace the colour changes of the single source candidates (ie. avoiding sources flagged as unresolved binaries) through the 4 \textit{K}-band magnitude bins by indicating where the bin members lie (with boxes) on a zoomed in version of the 2-colour diagram (shown in Figure \ref{fig:cc_emp}). The boxes were constructed such that single candidate members from each of the \textit{K}-band magnitude bins are all contained within the respective box dimensions in the diagram.
\par
Bin A, which is populated by candidates above the M dwarf gap, appears to be slightly bluer in (\textit{J-H}) than the NEXTGEN model colours, lying closer to the DUSTY model colours. We expect no dust condensation at these T$_{eff}$s, so this difference probably results from incompleteness in the NEXTGEN H$_2$O opacities (eg. see \citealt{jones03}). Bin B is slightly redder in \textit{H-K} than Bin A, but has similar \textit{J-H} colours. This colour change could be caused by the addition of new grain opacities, consistent with the appearance of atmospheric dust across the M dwarf gap. However, note that no colour changes were observed across the M dwarf gap in the Pleiades (see P03). If the M dwarf gap does result from the first appearance of atmospheric dust grains, then they must have different effects for different surface gravities. This would not be unexpected since we would expect grain growth and rain-out to be affected by both surface gravity and pressure. The fainter bins (C and D) have colours consistent with the Dusty models.
\section{THE MASSES OF LOW MASS CANDIDATES}
We have estimated the masses of all single Praesepe candidates that appear in our \textit{candidate sequence region} (see Figure 4). The exception to this is \textit{Iz}117, which was identified in $\S$\ref{sec:cc} as a likely non-member. Masses were estimated using the NEXTGEN and Dusty models (with (m-M)$_0$=6.16) for isochrone ages of 0.5 and 1 Gyrs, and solar metallicity. \citet{mart_2000c} note that mass estimates from \textit{J-} band magnitudes are the least sensitive to the choice of model, either NEXTGEN or DUSTY. Therefore, we choose to derive masses from this passband where possible. Where \textit{J-} band photometry was absent, masses were estimated from the \textit{K-} band magnitudes. 
\par
Linear interpolation was used in between the isochrone mass points. The DUSTY mass estimates are slightly higher than the NEXTGEN ones, and, rather more obviously, the younger mass estimates are lower than the older ones. The mass estimates are given in Table \ref{tab:photom_red}. Estimates taken from the \textit{K-} band are slightly higher than if taken from the \textit{J-} band.
\par
It can be seen from the table that we have identified at least 15 single VLMS candidates (with masses $\geq$ 0.075 M$_\odot$). The five faintest candidates could be sub-stellar or stellar depending on the age or model we adopt. For example, four of these are substellar if a 0.5 Gyr cluster age is assumed, regardless of which model is used.
\section{PRAESEPE BINARIES}
\subsection{Binary mass ratios}
\label{sec:bin_mf} 
In Figure \ref{fig:cm_k_emp} we have over-plotted 6 binary tracks, which we use to estimate mass ratios of our Praesepe binaries. We calculated these tracks by combining single star points from the cluster sequence. Each binary track begins at some single star point, and the affect of adding an unresolved companion is simulated by combining the photometric brightness of this single star point with other single star points from lower mass points in the sequence (beginning with the lowest mass point on the sequence, and then increasing the companion mass until an equal mass binary is reached, 0.75 magnitudes above the original single star point).
\par
When estimating binary mass ratios (q), one must account for the cluster depth which can make sources appear brighter without the need for a binary companion, as well as photometric uncertainty. The tidal radius of Praesepe is 12.1 pc \citep{holland} giving a depth effect $\sim$ $\pm$ 0.15 magnitudes. Photometric uncertainty is typically $\pm$0.2 in \textit{I-K}. Mass ratios were estimated by comparing unresolved binary candidates to the binary tracks in the CMD, and allowing for both these forms of uncertainty. The results are tabulated in Table \ref{tab:bin_mass}.  
\begin{table}
\caption{Binary mass ratios assuming a cluster depth of $\pm$ 0.15$^m$ using 0.5 Gyr and 1.0 Gyr NEXTGEN and Dusty models.  } \label{tab:bin_mass}
\begin{tabular}{|l||l|l||l|l|}
\hline
 &\multicolumn{2}{c|}{q (0.5 Gyr)}&\multicolumn{2}{c|}{q (1.0 Gyr)}\\
 &NG&DUSTY&NG&DUSTY\\
\hline\hline

RIZ2 & 0.8-1.0  & 0.8-1.0 &   0.9-1.0  & 0.9-1.0 \\
RIZ21 & 0.6-1.0 & 0.5-1.0 &0.6-1.0 & 0.6-1.0  \\
RIZ18 & 0.7-1.0 & 0.6-1.0 & 0.6-1.0 & 0.6-1.0 \\
IZ126 & 0.4-1.0 & 0.4-1.0 &0.5-1.0 & 0.5-1.0  \\
WFC169 & 0.7-1.0 & 0.7-1.0 & 0.8-1.0 & 0.7-1.0  \\
WFC118 & 0.6-1.0 & 0.6-1.0 & 0.7-1.0 & 0.6-1.0  \\
WFC88 & 0.7-1.0 & 0.7-1.0  &0.7-1.0 & 0.6-1.0   \\
WFC60 & 0.8-1.0 & 0.7-1.0   & 0.8-1.0 & 0.8-1.0 \\ 
\hline
\end{tabular}
\end{table}
\par
The q values range from 0.4--1.0. This represents a rather larger range of q values than was found for Pleiades BDs by P03 (q=0.7--1.0). However, the photometric uncertainties are slightly larger for the Praesepe candidates, and the very low mass unresolved binary populations identified in the two clusters cannot be shown to be significantly different (in their q range) by this analysis.
\subsection{The binary fraction}
P03 determined binary fractions (BFs; defined as the number of binary systems divided by the total number of systems in some photometric range) for Praesepe in 4 colour bins for 1 $\leq$ \textit{I-K} $\leq$ 3.6, corresponding to an approximate mass range 1.0 - 0.09 M$_\odot$. We extend this to lower masses by considering two redder bins from \textit{I-K} =3.3-3.9 and 3.9-4.6. For the first bin the \textit{Iz} survey is not complete over the full single star sequence. Therefore, to avoid underestimating the relative number of single star sources we only considered \textit{Riz} and WFC candidates. With 5 unresolved binary and 11 single source candidates, this bin has a BF$\sim$31\%. For the second bin, the \textit{Riz} survey is not complete over the full single star sequence. However, the \textit{Riz} candidates are all contained within the WFC survey area, and would have been discovered by this survey had they not already been identified. We therefore considered all sources in this bin, resulting in 3 unresolved binary and 7 single source candidates, or a BF$\sim$30\%. These results are summarised in Table \ref{tab:bf}, where uncertainties have been calculated assuming binomial statistics \citep{burg2003}. Note that the low number statistics give large uncertainties, and that the BF for the redder bin is a lot more uncertain even than this, since several of these candidates could be non-members (see Sections 5.4 and 10.4).
\begin{table}
\caption{Praesepe binary fractions. Mass and $q$ ranges estimated using Tables 2 and A3.} \label{tab:bf}
\begin{tabular}{|c|c|c|c|c|c|c|}
\hline

&  & \multicolumn{2}{c|} {MASS (0.001 M$_\odot$)} \\ \textit{I-K} range  &\multicolumn{2}{c|}{NEXTGEN}&\multicolumn{2}{c|}{DUSTY} & q & BF\\
\cline{2-5}
 &0.5&1.0&0.5&1.0\\
\hline\hline
3.3-3.9 &  ${89}{\pm 9}$  &${90}{\pm5}$  & ${70}{\pm5}$ & ${78}{\pm4}$ & 0.5-1.0 &${31}^{+13}_{-9}$ \\
3.9-4.6 &  ${73}{\pm 7}$  &${81}{\pm5}$  & ${63}{\pm4}$ & ${74}{\pm5}$ & 0.4-1.0 &${30}^{+17}_{-10}$ \\
\hline
\end{tabular} 
\end{table}
\par
\begin{figure}
\includegraphics[width=84mm]{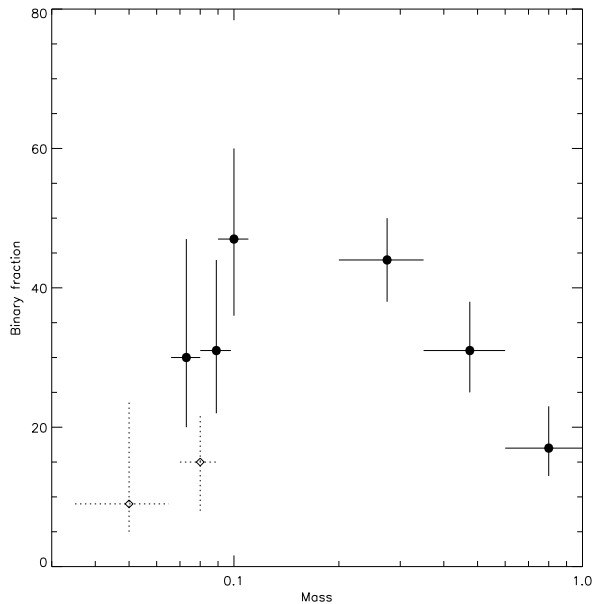}
\caption{Praesepe binary fractions. Two points from the literature are displayed as open diamonds.} \label{fig:bf} 
\end{figure}

The Praesepe BFs are displayed in Figure \ref{fig:bf}, with the 4 higher mass BFs from P03 over-plotted. Also shown are the BFs from \citet{close} and \citet{burg2003} for VLMS and BD binaries in the field with separations $>$1AU. Although our lowest mass point is too uncertain to accurately constrain the BF, the higher mass point shows that the BF is decreasing as one goes below 0.1M$_{\odot}$. Even if our expected level of contamination ($\sim$3 late M dwarfs) removes only single star candidates (thus pushing up the BF), the BF would only go up to 39$^{+14}_{-12}$\%, which would still suggest a BF decrease. This decreasing BF trend appears to be consistent with the lower BFs seen for field sources. P03 however, observed an increasing unresolved BF into the Pleiades BD regime, reaching 50$^{+10}_{-11}$ for 0.07M$_{\odot}$. This suggests that different clusters could have different very low mass BFs. Such differences could result from different cluster environments, and in this context the distribution of different cluster BFs could provide an important observational test of VLMS and BD formation theories. However, before any firm conclusions are drawn, it is clearly important to improve the robustness of the measured cluster BFs by confirming (or not) cluster membership from proper motions.
\section{LUMINOSITY AND MASS FUNCTION}
\label{sec:mf}
\begin{figure}
\includegraphics[width=84mm]{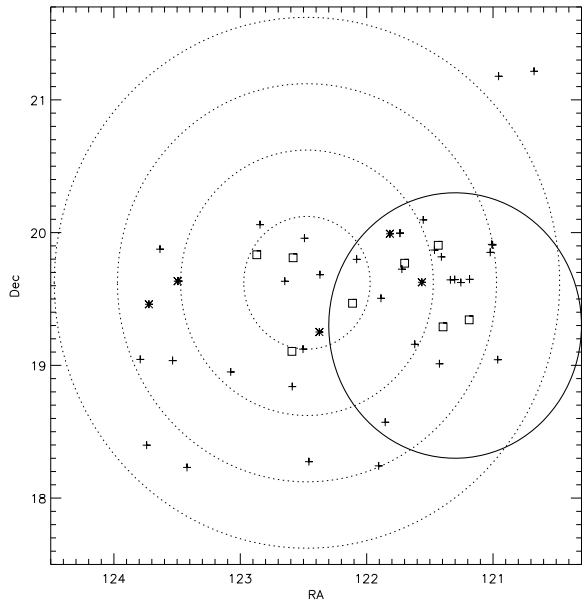}
\caption{Sky plot showing the positions of single candidates fainter than \textit{K-}=15.0. Crosses, squares and asterisks represent single candidates in the magnitude bins \textit{K=}15--15.8, 15.8--16.3 and 16.3--17.0 respectively. Annuli of radius 0.5$^o$, 1.0$^o$, 1.5$^o$ and 2.0$^o$ from the centre of the cluster are also shown as dotted lines. The position of the excluded subcluster is represented by the solid line.} \label{fig:numb_dens} 
\end{figure}

We divided our cluster members into 3 magnitude ranges covering \textit{K=}15--15.8, 15.8--16.3 and 16.3--17. For each magnitude range, cluster candidates were counted in 0.5 degree rings from the centre, out to a radius of 2.5 degrees for the brightest bin, and out to 2.0 degrees for the 2 fainter bins. We did not consider candidates found in the region of the sub-cluster identified by \citet{holland}, 3 pc from the centre of Praesepe. This sub-cluster could be a smaller older open cluster that has collided with Praesepe, and we avoided candidates within 1 degree of the sub-cluster centre. The positions of candidates in the magnitude bins \textit{K=}15--15.8, 15.8--16.3 and 16.3--17.0 are shown in figure \ref{fig:numb_dens} as  crosses, squares and asterisks respectively. It can be seen from the figure that there is an over-density of faint objects in the sub-cluster region. It is clearly important to isolate (and ignore) this region when measuring the LF and MFs of the main cluster, since the sub-cluster is thought to be older than the rest of Praesepe \citep{holland}, and it would otherwise be possible for faint low-mass sub-cluster stars to masquerade as Praesepe BDs.
\par
Ring counts were then converted into surface densities by dividing by the appropriate survey areas within each ring (ignoring the sub-cluster region), and expected levels of field star contamination (see Section \ref{sec:contamination}) were subtracted off the surface density profiles assuming a uniform spatial distribution over the cluster. We then assumed that the cluster could be represented by a King surface density distribution \citep{king} (as has been done previously by \citealt{pin_ple}, \citealt{jame} and \citealt{holland}). This function takes the form:
\begin{equation} \label{eq:king_profile}
f_{s}= k \left \{ \frac{1}{\sqrt{1+(r/r_{c})^2}} - \frac{1}{\sqrt{1+(r_{t}/r_{c})^2}}   \right \} ^2
\end{equation}
Here, $f_s$ is the surface density, r the radius from the cluster centre, \textit{k} a normalization constant and $r_c$ and $r_t$ the core and tidal radii respectively. Assuming $r_t$ = 12.1 pc (\citealt{holland}), and using Poisson uncertainties for the surface density profiles, we obtained the best fit value of $r_c$ using the whole sample, and found $r_c$=7.56 pc. This was then assumed to be constant for each magnitude bin, so we proceeded to minimize the $\chi^2$ statistic of the King profile to the 3 observed density distributions. We then determined the total number of cluster stars in each bin using the surface integral (out to $r_t$) of equation \ref{eq:king_profile}; 

\begin{figure}
\includegraphics[width=84mm]{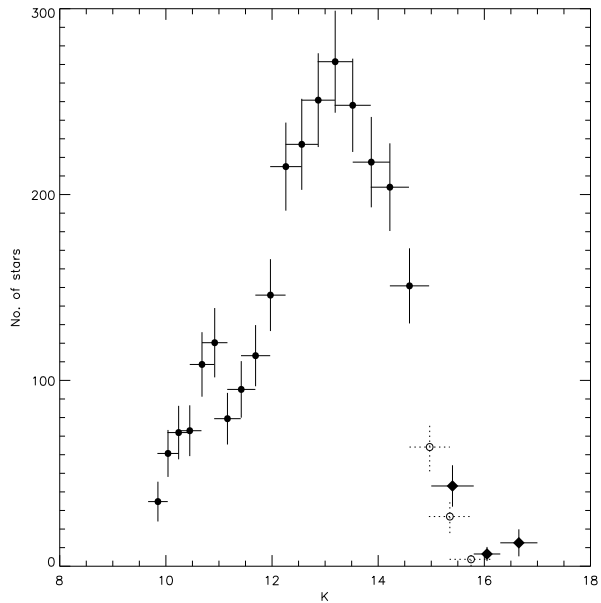}
\caption{Praesepe \textit{K-}band luminosity function. Diamonds are points derived in this work. Circles are from the HSHJ survey, where the last three HSHJ points are lower limits. } \label{fig:lum_func}
\end{figure}

\begin{equation}
n=\pi r_c^2K \left \{  ln(1+x_t) - \frac{(3 \sqrt{1+x_t} -1)(\sqrt{1+x_t}-1)}{1+x_t} \right \}
\end{equation}
where $x_t = (r_t/r_c)^2$. Our results are summarised in Table \ref{tab:lf_mf}. The resulting luminosity function ($\Phi_K$) is shown in figure \ref{fig:lum_func} as filled diamonds. Also shown is the luminosity function from HSHJa (converted into $\Phi_{K}$), although note that the faintest three HSHJ points (open circles) are lower limits due to photometric incompleteness.

The mass function then follows directly from the luminosity function:
\begin{equation}
N(m) = \Phi_K \frac{dK} {dm}.
\end{equation}
We estimated $\frac{dK}{dm}$ for our 3 magnitude ranges using both the 0.5 Gyr and 1.0 Gyr NEXTGEN and DUSTY isochrones to define the mass limits of each bin, and derived the cluster mass function per 0.1M$_{\odot}$ interval (see Table \ref{tab:lf_mf}).

\begin{table*}
\caption{Luminosity and mass function bins with derived values as discussed in the text.}
\label{tab:lf_mf} 
\begin{tabular}{|c|c|c|c|c|c|c|c|c|c|c|c|c|}
\hline
&&&&&&\multicolumn{2}{c|}{Mass Range (0.001 M$_{\odot}$)} & 
\multicolumn{2}{c|}{N(m) /0.1 M$_\odot$}\\
\multicolumn{1}{|c|}{K Range} & 
\multicolumn{1}{c}{Surveys}& 
\multicolumn{1}{c}{N$^{o.}$}&
\multicolumn{1}{c}{Contam.} & 
\multicolumn{1}{c}{N(r$_t$)} & 
\multicolumn{1}{c}{$\Phi_{K}$} &
\multicolumn{1}{c}{0.5 Gyr} &
\multicolumn{1}{|c|}{1.0 Gyr} &
\multicolumn{1}{|c|}{0.5 Gyr} &
\multicolumn{1}{|c|}{1.0 Gyr} 
\\
\hline
15.0--15.8 &  Riz,Iz,M98,WFC & 15 & 3.0  & 34.6$\pm$8.9 & 43.2$\pm$11.2 & 
87--123 & 91--122 & 96.0$\pm$24.8 & 111.5$\pm$28.8 \\
15.8--16.3 &  Riz,M98,WFC & 3 & 2.5  & 3.3$\pm$1.9& 6.6$\pm$3.8& 
74--93 & 84--96 & 17.3$\pm$10.0 & 27.4$\pm$15.8 \\
16.3--17.0 &  M98,WFC & 3 & 2.0  & 8.8$\pm$5.1 & 12.6$\pm$7.3 & 
61--78 & 74--86 & 51.9$\pm$29.9 & 73.5$\pm$42.4 \\
\hline
\end{tabular} 
\end{table*}

\begin{figure}
\includegraphics[width=84mm]{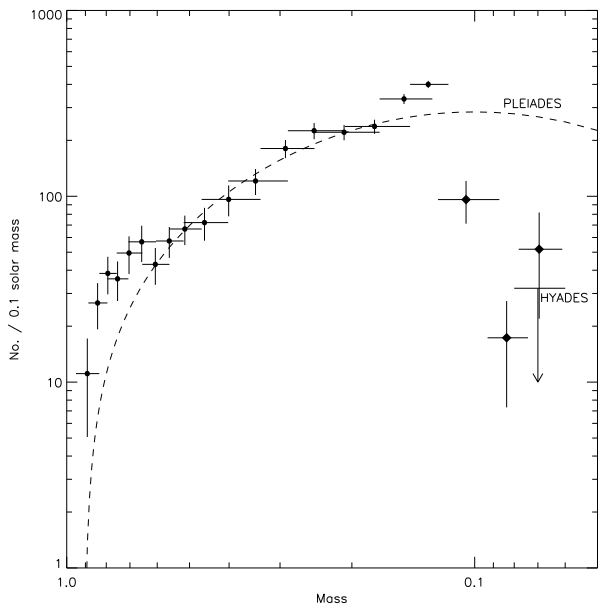}
\caption{Praesepe mass function. Symbols are same as for figure \ref{fig:lum_func}. The log-normal Pleiades MF best-fit from \citet{adams01} is shown as a dotted line, normalized  at 0.5 M$_\odot$. Also shown is a 0.06--0.08 M$_\odot$ MF upper limit for the Hyades cluster taken from \citet{dob_hyades}.  } \label{fig:mf}
\end{figure}
The log-normal Pleiades mass function as derived by \citet{adams01} is displayed on Figure \ref{fig:mf} as a dashed line, and is normalised to the Praesepe function at 0.5 M$_\odot$. An upper limit Hyades mass function point from \citet{dob_hyades} in the range 0.08-0.06 M$_\odot$ is also shown for comparison. 
\par
It can be seen from Figure 11 that both the Praesepe mass function derived here and the Hyades mass function falls below that of the Pleiades at $\sim$0.1M$_{\odot}$. The Hyades and Praesepe are thought to be of a similar age ($\sim$ 0.6 Gyr), compared to the younger 0.13 Gyr Pleiades cluster \citep{burke}. This age gap has lead \citet{dob_hyades} to suggest that the paucity of low-mass Hyads is caused by dynamical evolution of the environment. In this scenario, the lowest-mass members of an association are, over time, preferentially ejected from the cluster -- a theory supported by the \textit{N}--body simulations of \citet{fuente_marcos} in which a 0.6 Gyr open cluster has retained at best only $\sim$ 17 percent of the BDs that were originally bound to the association. Evidently, our Praesepe mass function is in strong agreement with this scenario.  
\par
Despite the qualitative similarities, however, between the mass functions of Praesepe and Hyades, there are to date no Hyad substellar candidates, in comparison to the Praesepe candidates proposed in this work. As a result, the Hyades low-mass MF upper limit is only marginally consistent with our lowest mass Praesepe point, which suggests that the Praesepe MF may be somewhat higher than that of the Hyades in this mass range. Clearly, differing dynamical processes would not be an issue here since both clusters are of a similar age and can be assumed therefore to be at a similar stage of evolution.  If this is a real feature of the MF, rather than as a consequence of low-number statistics,  we speculate that this difference can perhaps be explained by a substellar mass function sensitivity to natal molecular cloud properties such as density, metallicity, temperature etc. Indeed, the simulations of \citet{elm2000} suggest that the shape of the MF below the observed 0.1 M$_\odot$ turnover is dependent upon the physical cloud properties, in contrast to the universal Salpeter IMF that characterises the intermediate to high mass domain. We stress, though, that it is imperative to increase the known substellar population in a variety of associations before this hypothesis can be fully tested, and that the possibility of a common substellar MF in both environments cannot be ruled out.

\section{ASTROMETRY}
\label{sec:astrometry}
We have mentioned, in previous sections, how important proper motions are in confirming cluster membership. The peculiar motion of Praesepe is large compared to that of typical field stars, and the intrinsic velocity dispersion of the cluster is small (HSHJb). With a base-line of a few years, it is thus possible to measure the proper motion of candidates with sufficient accuracy to separate true cluster members from field star contamination. In this section we describe how we have measured proper motions for a subset of the Praesepe candidates we consider in this work.
\subsection{Epoch data}
First epoch data for the astrometric analysis of Praesepe RIZ candidates was taken in 1993 using the 2.5 m Isaac Newton Telescope (INT), providing a baseline of $\sim$ 8.9 years. For Roque PR1, we use the original image acquired in February 1996, also taken with the INT, with a baseline of $\sim$ 5.9 years. We use our WFC images as 2nd epoch data. The images in which RIZ 21 and 23 were recovered were acquired on the third night of the observing run, which could not be photometrically calibrated due to poor conditions. Consequently, these two candidates are not plotted in Figure \ref{fig:iz_cm}. However, we do not expect this to affect the astrometric measurements. For all images, we use only the \textit{I} band for the astrometry. 

\subsection{Measurements}
The process of deriving the astrometry of the cluster candidates is as follows. The pixel positions of the candidates on both epoch images ($x_{1st}$,$y_{1st}$),($x_{2nd}$,$y_{2nd}$) are first accurately determined using the IRAF:CENTRE routine. The positions of presumed background point sources were also measured in both epochs, to be used as astrometric references. IRAF:XYXYMATCH was used to calculate the transformation function which maps the two epoch images. GEOMAP and GEOXYTRAN were then used to transpose the second epoch coordinates of the candidates into the first epoch coordinate system. Any reference sources lying more than a Gaussian 3 $\sigma$  from the median of absolute deviation were rejected and the transform redetermined.

\par
The relative proper motions along the x and y axes $\mu_{x,y}$ of the CCD then follows from the pixel displacement:
\begin{equation}
\mu_{X} = \frac{(\Delta X)}{\Delta t} = \frac{X_{2nd}-X_{1st}}{\Delta t }
\end{equation} 
\begin{equation}
\mu_{Y} = \frac{(\Delta Y)}{\Delta t} = \frac{Y_{2nd}-Y_{1st}}{\Delta t }
\end{equation} 
The uncertainties of the transformations $\sigma_{x,y}$ were found to be $\approx$ 0.2 pixels, corresponding to proper motion errors of  $\approx$ 10-15 masyr$^{-1}$, in both RA and Dec direction.

\subsection{Membership probabilities}
The method of establishing membership probabilities is that outlined by \citet{sand}, and utilizes the following likelihood function:

\begin{equation}
L=\prod_{i=1}^{N} \Phi (\mu _{x},\mu _{y})
\end{equation}

where 

\begin{equation}
\Phi (\mu _{x},\mu _{y}) = \frac {n_c}{2 \pi \sigma ^{2}} \Phi _{c} (\mu _{x},\mu _{y}) + \frac {1-n_c}{2 \pi \Sigma _{x} \Sigma _{y} (1- \rho ^{2})} \Phi _{f} (\mu _{x},\mu _{y})
\end{equation}

here, $n_c$ is the normalized number of cluster stars, and $\rho$ is a correlation coefficient. Following the method of \citet{wang}, $\sigma$ follows from the proper motion uncertainties $\sigma _{x,y}$ and an intrinsic variance $\sigma _{0}$ :  
\begin{equation}
\sigma ^ 2 = \sigma _{0} ^2 + \sigma _{x,y} ^2 
\end{equation}
\par
The distribution functions of the cluster and field $\Phi_{c},\Phi_{f}$ are evaluated as:

\begin{equation}
\Phi _{c} (\mu _{x},\mu _{y}) = exp \left \{ - \frac {1}{2} \left[{ \left(\frac {\mu _{x}- \mu _{xc}} {\sigma} \right)}^{2} + \left(\frac {\mu _{y}- \mu _{yc}} {\sigma} \right) ^{2} \right] \right\}
\end{equation} 

and

\begin{multline}
\Phi _{f} (\mu _{x},\mu _{y}) = exp \left \{ - \frac {1}{2(1- \rho ^{2})} \left [ \left( \frac {\mu _{x} - \mu _{xf}}{\Sigma _{x}}  \right) ^2  \right . \right . \\  
\left . \left.   - \frac {2 \rho ( \mu _{x} - \mu _{yf}) ( \mu _{y} - \mu _{yf})}{ \Sigma _{x} \Sigma 
 _{y}}+ \left ( \frac { \mu _{y} - \mu _{yf}}{ \Sigma _{y}} \right ) ^2 \right ] \right \} 
\end{multline}

here, $\sigma$ is the proper motion dispersion, $\mu _{xc}$,$\mu _{yc}$ are the cluster mean proper motions, and $\mu _{xf}$,$\mu _{yf}$ are the field mean proper motions with standard deviations $\Sigma _x$ and $\Sigma _y$.
\par
The free parameters were found by \citet{wang} using a conjugate gradient method using a large Praesepe census, and are reproduced in Table \ref{tab:wang_par}. 

\begin{table}
\caption{Praesepe maximum likelihood parameters} \label{tab:wang_par} 
\
\begin{tabular}{cc}
\hline
\multicolumn{1}{c}{parameter} & \multicolumn{1}{c}{value}  \\
\hline
$\mu _{xc}$ & -32.88 $\pm$ 0.09 mas yr$^{-1}$ \\
$\mu _{yc}$ & -15.87 $\pm$ 0.09 mas yr$^{-1}$ \\
$\mu _{xf}$ & 0.23 $\pm$ 0.08 mas yr$^{-1}$ \\
$\mu _{yf}$ & -7.18 $\pm$ 0.26 mas yr$^{-1}$ \\
$\sigma_{0}$ & 1.45 $\pm$ 0.09 mas yr$^{-1}$ \\
$\rho$ & 0.002 $\pm$ 0.016 \\
\hline
\end{tabular} 
\end{table}

\subsection{Astrometric results}
\begin{figure}
\includegraphics[width=84mm]{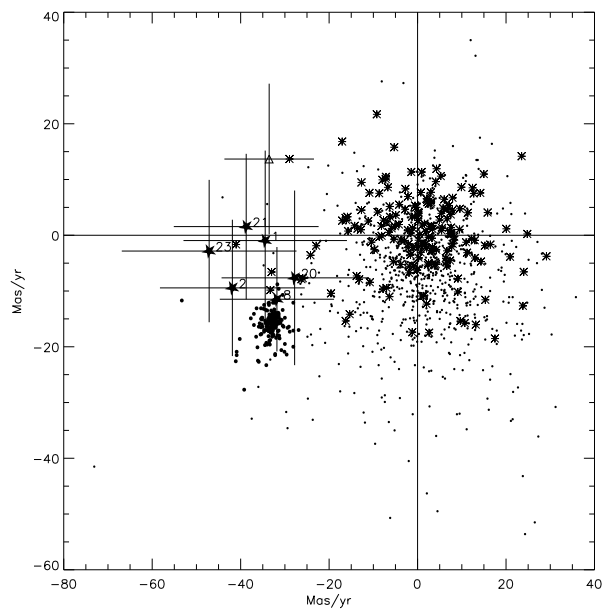}
\caption{Proper motion vector point diagram for objects located in Praesepe fields. Asterisks are the sources used as astrometric references. Filled stars show the RIZ candidates, and the triangle is RPr1. Field and cluster members (P $>$ 70 percent) assigned in \citet{wang} are shown as small and large filled circles respectively. } \label{fig:vpd} 
\end{figure}

Proper motions and membership probabilities of RIZ objects from P97 that have JHK colours consistent with cluster memberships are shown in Table \ref{tab:ast_res} and illustrated in a vector point diagram (VPD) in Figure \ref{fig:vpd} as filled stars with error bars. We were unable to measure the proper motion of RIZ 4,11,18 and 24 due to insufficient shared spatial coverage between epochs, which would have resulted in unreliable transformations. 
\par
Proper motions from \citet{wang} are overplotted as small circles, or large circles where membership probabilities were evaluated to be greater than 70 percent. These help to clarify the cluster position at $\mu_{x,y} \approx$ -33,-16 $mas yr^{-1}$. All stars that were used as references for the astrometric transformation are also shown on the VPD as asterisks. As expected, they cluster around the origin of the diagram, since they were selected on the basis of having a negligible motion.

\par
All of the RIZ candidates have an evaluated membership probability greater than 70 percent, and are therefore consistant with cluster membership within the astrometric uncertainties. Our astrometric analysis of Roque Pr1, shown as a triangle on the VPD, suggests that this object is most likely a non-cluster member. RPr1 is more likely to be a field star in the line of sight to the cluster.   

\begin{table}
\caption{Astrometric results. $\ddag$ is also WFC91.} \label{tab:ast_res} 
\
\begin{tabular}{ccccc}
\hline
\multicolumn{1}{c}{Object} & 
\multicolumn{1}{c}{$\mu_{x}$ mas yr$^{-1}$} & 
\multicolumn{1}{c}{$\mu_{y}$ mas yr$^{-1}$} & 
\multicolumn{1}{c}{Prob} & 
\multicolumn{1}{c}{mem}  \\
\hline
RIZ1$^\ddag$ & -34.45 $\pm$ 18.48  & -0.95 $\pm$ 16.14  & 86.72 & Y\\
RIZ2 & -41.87 $\pm$ 16.37  & -9.44 $\pm$ 12.23 &  94.33 & Y\\
RIZ8 & -31.80  $\pm$ 12.92 & -11.48$\pm$ 9.37  & 97.87 & Y\\
RIZ20 & -27.80 $\pm$ 16.52  & -7.64 $\pm$ 15.66 & 86.56 & Y\\
RIZ21 & -38.75 $\pm$ 16.37  & -1.56 $\pm$ 13.05  & 72.11 & Y\\
RIZ23 & -47.12 $\pm$ 19.75  & -2.83 $\pm$ 12.77  & 75.41 & Y\\
RPr1 & -33.55 $\pm$ 10.13    & 13.67 $\pm$ 13.52   & 51.24 & N?\\
\hline
\end{tabular} 
\end{table}

\section{CONCLUSIONS \& FUTURE WORK}

In this paper, we have presented the results of a deep optical survey (I$\sim$21.3) and NIR follow-up of 2.6 square degrees of the Praesepe open cluster. We have used our census of cool red members to place constraints on the bottom end of the cluster mass function. Our results indicate that the dynamical evolution of the cluster is likely leading to the evaporation of the lowest mass members. However, the relative depletion of the very-low mass stellar and substellar members may be less than that reported for the similarly aged Hyades cluster, suggesting that perhaps the cluster MF is sensitive to natal molecular cloud properties. 
\par
To test the robustness of these arguments, it is imperative to ascertain beyond doubt the validity of the candidates' claims to  cluster membership. We have done this using proper motion measurements for several low-mass Praesepe candidates, but our priority is to continue this for \textit{all}. Only then can we probe with confidence the substellar MF. Furthermore, a \textit{bona-fide} census would strengthen the  comparison between low-mass binary fractions in this cluster compared to earlier clusters such as the Pleiades. Our results suggest a difference between the two environments, but its cause or  quantitative nature cannot be elaborated upon without further census expansion. 
\par
The photometric sample we have compiled have revealed a paucity of candidates close to the region where the main 
sequence begins to deviate away from the NEXTGEN track. This feature has been observed in younger 
regions too, and would appear to originate because of a steepening in the mass-luminosity relation at 
late M spectral type. We propose that further measurements, both photometric (e.g \textit{L-} band) 
and spectroscopic, be made of candidates which are located just above and below the gap in both this 
cluster and others to isolate the underlying physics, and to test the sensitivity of dust formation 
in the atmosphere of early cluster L dwarfs to the fundamental properties of their association.  
\par
Finally, to further the results presented in this work it is of utmost importance to increase 
the known number of VLMS and BDs in young SFRs and open clusters. In the near future, this need will 
primarily be met by the advent of automated deep surveys conducted over wide areas. One such effort 
is UKIDSS, the collective name for a series of large-scale IR surveys using the WFCAM instrument on the UK Infrared Telescope. The Galactic Cluster Survey (GCS) in particular will survey 10 associations over the next few years on an unprecedented scale, uncovering 
many more cluster BDs that will go a long way to addressing some of the issues raised in this 
work.

\section{Acknowledgments}
RJC acknowledges financial support from The Particle Physics and Astronomy Research Council (PPARC) of the UK. PDD is sponsored by a PPARC postdoctoral grant. This publication makes use of data products from the Two Micron All Sky Survey, which is a joint project of the University of Massachusetts and the Infrared Processing and Analysis Center/California Institute of Technology, funded by the National Aeronautics and Space Administration and the National Science Foundation. The United Kingdom Infrared Telescope is operated by the Joint Astronomy Centre on behalf of the UK Particle Physics and Astronomy Research Council. Some of the data reported here were obtained as part of the UKIRT Service Programme.
\bibliographystyle{mn2e}
\bibliography{refs}

\begin{thebibliography}{}

\bibitem[\protect\citeauthoryear{{Adams}, {Stauffer}, {Monet}, {Skrutskie} \&
  {Beichman}}{{Adams} et~al.}{2001}]{adams01}
{Adams} J.~D.,  {Stauffer} J.~R.,  {Monet} D.~G.,  {Skrutskie} M.~F.,
  {Beichman} C.~A.,  2001, AJ, 121, 2053

\bibitem[\protect\citeauthoryear{{Adams}, {Stauffer}, {Skrutskie}, {Monet},
  {Portegies Zwart}, {Janes} \& {Beichman}}{{Adams} et~al.}{2002}]{adams02}
{Adams} J.~D.,  {Stauffer} J.~R.,  {Skrutskie} M.~F.,  {Monet} D.~G.,
  {Portegies Zwart} S.~F.,  {Janes} K.~A.,    {Beichman} C.~A.,  2002, AJ, 124,
  1570

\bibitem[\protect\citeauthoryear{{Baraffe}, {Chabrier}, {Allard} \&
  {Hauschildt}}{{Baraffe} et~al.}{1998}]{bar98}
{Baraffe} I.,  {Chabrier} G.,  {Allard} F.,    {Hauschildt} P.~H.,  1998, A\&A,
  337, 403

\bibitem[\protect\citeauthoryear{{Bessell}}{{Bessell}}{1986}]{bessell}
{Bessell} M.~S.,  1986, PASP, 98, 1303

\bibitem[\protect\citeauthoryear{{Bouvier}, {Stauffer}, {Martin}, {Barrado y
  Navascues}, {Wallace} \& {Bejar}}{{Bouvier} et~al.}{1998}]{bouv_98}
{Bouvier} J.,  {Stauffer} J.~R.,  {Martin} E.~L.,  {Barrado y Navascues} D.,
  {Wallace} B.,    {Bejar} V.~J.~S.,  1998, A\&A, 336, 490

\bibitem[\protect\citeauthoryear{{Bridger}, {Wright}, {Economou}, {Tan},
  {Currie}, {Pickup}, {Adamson}, {Rees}, {Purves} \& {Kackley}}{{Bridger}
  et~al.}{2000}]{bridger}
{Bridger} A.,  {Wright} G.~S.,  {Economou} F.,  {Tan} M.,  {Currie} M.~J.,
  {Pickup} D.~A.,  {Adamson} A.~J.,  {Rees} N.~P.,  {Purves} M.,    {Kackley}
  R.,  2000, in Proc. SPIE Vol. 4009, p. 227-238, Advanced Telescope and
  Instrumentation Control Software, Hilton Lewis; Ed. {ORAC: a modern observing
  system for UKIRT}.
pp 227--238

\bibitem[\protect\citeauthoryear{{Burgasser}, {Kirkpatrick}, {Reid}, {Brown},
  {Miskey} \& {Gizis}}{{Burgasser} et~al.}{2003}]{burg2003}
{Burgasser} A.~J.,  {Kirkpatrick} J.~D.,  {Reid} I.~N.,  {Brown} M.~E.,
  {Miskey} C.~L.,    {Gizis} J.~E.,  2003, ApJ, 586, 512

\bibitem[\protect\citeauthoryear{{Burke}, {Pinsonneault} \& {Sills}}{{Burke}
  et~al.}{2004}]{burke}
{Burke} C.~J.,  {Pinsonneault} M.~H.,    {Sills} A.,  2004, ApJ, 604, 272

\bibitem[\protect\citeauthoryear{{Carpenter}}{{Carpenter}}{2001}]{carp}
{Carpenter} J.~M.,  2001, AJ, 121, 2851

\bibitem[\protect\citeauthoryear{{Chabrier}, {Baraffe}, {Allard} \&
  {Hauschildt}}{{Chabrier} et~al.}{2000}]{chab2000}
{Chabrier} G.,  {Baraffe} I.,  {Allard} F.,    {Hauschildt} P.,  2000, ApJ,
  542, 464

\bibitem[\protect\citeauthoryear{{Close}, {Siegler}, {Freed} \&
  {Biller}}{{Close} et~al.}{2003}]{close}
{Close} L.~M.,  {Siegler} N.,  {Freed} M.,    {Biller} B.,  2003, ApJ, 587, 407

\bibitem[\protect\citeauthoryear{{Cooper}, {Sudarsky}, {Milsom}, {Lunine} \&
  {Burrows}}{{Cooper} et~al.}{2003}]{cooper}
{Cooper} C.~S.,  {Sudarsky} D.,  {Milsom} J.~A.,  {Lunine} J.~I.,    {Burrows}
  A.,  2003, ApJ, 586, 1320

\bibitem[\protect\citeauthoryear{{Cossburn}, {Hodgkin}, {Jameson} \&
  {Pinfield}}{{Cossburn} et~al.}{1998}]{cossburn}
{Cossburn} M.~R.,  {Hodgkin} S.~T.,  {Jameson} R.~F.,    {Pinfield} D.~J.,
  1998, in ASP Conf. Ser. 154: Cool Stars, Stellar Systems, and the Sun
  {Defining a CCD Infrared Colour-Temperature Relationship for Low-Mass Stars}.
pp 1854--+

\bibitem[\protect\citeauthoryear{{Crawford} \& {Barnes}}{{Crawford} \&
  {Barnes}}{1969}]{crawford69}
{Crawford} D.~L.,  {Barnes} J.~V.,  1969, AJ, 74, 818

\bibitem[\protect\citeauthoryear{{Cruz}, {Reid}, {Liebert}, {Kirkpatrick} \&
  {Lowrance}}{{Cruz} et~al.}{2003}]{cruz}
{Cruz} K.~L.,  {Reid} I.~N.,  {Liebert} J.,  {Kirkpatrick} J.~D.,    {Lowrance}
  P.~J.,  2003, AJ, 126, 2421

\bibitem[\protect\citeauthoryear{{Dahn}, {Harris}, {Vrba}, {Guetter},
  {Canzian}, {Henden}, {Levine}, {Luginbuhl}, {Monet}, {Monet}, {Pier},
  {Stone}, {Walker}, {Burgasser}, {Gizis}, {Kirkpatrick}, {Liebert} \&
  {Reid}}{{Dahn} et~al.}{2002}]{dahn}
{Dahn} C.~C.,  {Harris} H.~C.,  {Vrba} F.~J.,  {Guetter} H.~H.,  {Canzian} B.,
  {Henden} A.~A.,  {Levine} S.~E.,  {Luginbuhl} C.~B.,  {Monet} A.~K.~B.,
  {Monet} D.~G.,  {Pier} J.~R.,  {Stone} R.~C.,  {Walker} R.~L.,  {Burgasser}
  A.~J.,  {Gizis} J.~E.,  {Kirkpatrick} J.~D.,  {Liebert} J.,    {Reid} I.~N.,
  2002, AJ, 124, 1170

\bibitem[\protect\citeauthoryear{{de la Fuente Marcos} \& {de la Fuente
  Marcos}}{{de la Fuente Marcos} \& {de la Fuente
  Marcos}}{2000}]{fuente_marcos}
{de la Fuente Marcos} R.,  {de la Fuente Marcos} C.,  2000, Ap\&SS, 271, 127

\bibitem[\protect\citeauthoryear{{Dobbie}, {Kenyon}, {Jameson} \&
  {Hodgkin}}{{Dobbie} et~al.}{2002}]{dobbie_iz}
{Dobbie} P.~D.,  {Kenyon} F.,  {Jameson} R.~F.,    {Hodgkin} S.~T.,  2002,
  MNRAS, 331, 445

\bibitem[\protect\citeauthoryear{{Dobbie}, {Kenyon}, {Jameson}, {Hodgkin},
  {Hambly} \& {Hawkins}}{{Dobbie} et~al.}{2002}]{dob_hyades}
{Dobbie} P.~D.,  {Kenyon} F.,  {Jameson} R.~F.,  {Hodgkin} S.~T.,  {Hambly}
  N.~C.,    {Hawkins} M.~R.~S.,  2002, MNRAS, 329, 543

\bibitem[\protect\citeauthoryear{{Dobbie}, {Kenyon}, {Jameson}, {Hodgkin},
  {Pinfield} \& {Osborne}}{{Dobbie} et~al.}{2002}]{dobb02}
{Dobbie} P.~D.,  {Kenyon} F.,  {Jameson} R.~F.,  {Hodgkin} S.~T.,  {Pinfield}
  D.~J.,    {Osborne} S.~L.,  2002, MNRAS, 335, 687

\bibitem[\protect\citeauthoryear{{Dobbie}, {Pinfield}, {Jameson} \&
  {Hodgkin}}{{Dobbie} et~al.}{2002}]{dobbie}
{Dobbie} P.~D.,  {Pinfield} D.~J.,  {Jameson} R.~F.,    {Hodgkin} S.~T.,  2002,
  MNRAS, 335, L79

\bibitem[\protect\citeauthoryear{{Elmegreen}}{{Elmegreen}}{2000}]{elm2000}
{Elmegreen} B.~G.,  2000, MNRAS, 311, L5

\bibitem[\protect\citeauthoryear{{Hambly}, {Steele}, {Hawkins} \&
  {Jameson}}{{Hambly} et~al.}{1995a}]{hshja}
{Hambly} N.~C.,  {Steele} I.~A.,  {Hawkins} M.~R.~S.,    {Jameson} R.~F.,
  1995a, MNRAS, 273, 505

\bibitem[\protect\citeauthoryear{{Hambly}, {Steele}, {Hawkins} \&
  {Jameson}}{{Hambly} et~al.}{1995b}]{hshjb}
{Hambly} N.~C.,  {Steele} I.~A.,  {Hawkins} M.~R.~S.,    {Jameson} R.~F.,
  1995b, A\&AS, 109, 29

\bibitem[\protect\citeauthoryear{{Hawarden}, {Leggett}, {Letawsky},
  {Ballantyne} \& {Casali}}{{Hawarden} et~al.}{2001}]{hawarden}
{Hawarden} T.~G.,  {Leggett} S.~K.,  {Letawsky} M.~B.,  {Ballantyne} D.~R.,
  {Casali} M.~M.,  2001, MNRAS, 325, 563

\bibitem[\protect\citeauthoryear{{Hodgkin}, {Pinfield}, {Jameson}, {Steele},
  {Cossburn} \& {Hambly}}{{Hodgkin} et~al.}{1999}]{hodge}
{Hodgkin} S.~T.,  {Pinfield} D.~J.,  {Jameson} R.~F.,  {Steele} I.~A.,
  {Cossburn} M.~R.,    {Hambly} N.~C.,  1999, MNRAS, 310, 87

\bibitem[\protect\citeauthoryear{{Holland}, {Jameson}, {Hodgkin}, {Davies} \&
  {Pinfield}}{{Holland} et~al.}{2000}]{holland}
{Holland} K.,  {Jameson} R.~F.,  {Hodgkin} S.,  {Davies} M.~B.,    {Pinfield}
  D.,  2000, MNRAS, 319, 956

\bibitem[\protect\citeauthoryear{{Jameson}, {Dobbie}, {Hodgkin} \&
  {Pinfield}}{{Jameson} et~al.}{2002}]{jame}
{Jameson} R.~F.,  {Dobbie} P.~D.,  {Hodgkin} S.~T.,    {Pinfield} D.~J.,  2002,
  MNRAS, 335, 853

\bibitem[\protect\citeauthoryear{{Jones}, {Pavlenko}, {Viti} \&
  {Tennyson}}{{Jones} et~al.}{2003}]{jones03}
{Jones} H.~R.~A.,  {Pavlenko} Y.,  {Viti} S.,    {Tennyson} J.,  2003, in The
  Future of Cool-Star Astrophysics: 12th Cambridge Workshop on Cool Stars,
  Stellar Systems, and the Sun (2001 July 30 - August 3), eds. A. Brown, G.M.
  Harper, and T.R. Ayres, (University of Colorado), 2003, p. 899-905. {A
  Comparison of Water Vapour Line Lists}.
pp 899--905

\bibitem[\protect\citeauthoryear{{Jones} \& {Tsuji}}{{Jones} \&
  {Tsuji}}{1997}]{jones_97}
{Jones} H.~R.~A.,  {Tsuji} T.,  1997, ApJL, 480, L39+

\bibitem[\protect\citeauthoryear{{King}}{{King}}{1962}]{king}
{King} I.,  1962, AJ, 67, 471

\bibitem[\protect\citeauthoryear{{Kirkpatrick}, {McGraw}, {Hess}, {Liebert} \&
  {McCarthy}}{{Kirkpatrick} et~al.}{1994}]{kirk_94}
{Kirkpatrick} J.~D.,  {McGraw} J.~T.,  {Hess} T.~R.,  {Liebert} J.,
  {McCarthy} D.~W.,  1994, ApJS, 94, 749

\bibitem[\protect\citeauthoryear{{Knapp}, {Leggett}, {Fan}, {Marley},
  {Geballe}, {Golimowski}, {Finkbeiner}, {Gunn}, {Hennawi}, {Ivezi{\' c}},
  {Lupton}, {Schlegel}, {Strauss} \& {Tsvetanov}}{{Knapp} et~al.}{2004}]{knapp}
{Knapp} G.~R.,  {Leggett} S.~K.,  {Fan} X.,  {Marley} M.~S.,  {Geballe} T.~R.,
  {Golimowski} D.~A.,  {Finkbeiner} D.,  {Gunn} J.~E.,  {Hennawi} J.,
  {Ivezi{\' c}} Z.,  {Lupton} R.~H.,  {Schlegel} D.~J.,  {Strauss} M.~A.,
  {Tsvetanov} Z.~I.,  2004, AJ, 127, 3553

\bibitem[\protect\citeauthoryear{{Landolt}}{{Landolt}}{1992}]{land}
{Landolt} A.~U.,  1992, AJ, 104, 340

\bibitem[\protect\citeauthoryear{{Leggett}, {Allard}, {Geballe}, {Hauschildt}
  \& {Schweitzer}}{{Leggett} et~al.}{2001}]{legg_01}
{Leggett} S.~K.,  {Allard} F.,  {Geballe} T.~R.,  {Hauschildt} P.~H.,
  {Schweitzer} A.,  2001, ApJ, 548, 908

\bibitem[\protect\citeauthoryear{{Leggett}, {Allard} \& {Hauschildt}}{{Leggett}
  et~al.}{1998}]{legg_98}
{Leggett} S.~K.,  {Allard} F.,    {Hauschildt} P.~H.,  1998, ApJ, 509, 836

\bibitem[\protect\citeauthoryear{{Leggett}, {Golimowski}, {Fan}, {Geballe},
  {Knapp}, {Brinkmann}, {Csabai}, {Gunn}, {Hawley}, {Henry}, {Hindsley},
  {Ivezi{\' c}}, {Lupton}, {Pier}, {Schneider}, {Smith}, {Strauss}, {Uomoto} \&
  {York}}{{Leggett} et~al.}{2002}]{leg02}
{Leggett} S.~K.,  {Golimowski} D.~A.,  {Fan} X.,  {Geballe} T.~R.,  {Knapp}
  G.~R.,  {Brinkmann} J.,  {Csabai} I.,  {Gunn} J.~E.,  {Hawley} S.~L.,
  {Henry} T.~J.,  {Hindsley} R.,  {Ivezi{\' c}} {\v Z}.,  {Lupton} R.~H.,
  {Pier} J.~R.,  {Schneider} D.~P.,  {Smith} J.~A.,  {Strauss} M.~A.,  {Uomoto}
  A.,    {York} D.~G.,  2002, ApJ, 564, 452

\bibitem[\protect\citeauthoryear{{Magazzu}, {Rebolo}, {Zapatero Osorio},
  {Martin} \& {Hodgkin}}{{Magazzu} et~al.}{1998}]{mag}
{Magazzu} A.,  {Rebolo} R.,  {Zapatero Osorio} M.~R.,  {Martin} E.~L.,
  {Hodgkin} S.~T.,  1998, ApJL, 497, L47+

\bibitem[\protect\citeauthoryear{{Mart{\'{\i}}n}, {Brandner}, {Bouvier},
  {Luhman}, {Stauffer}, {Basri}, {Zapatero Osorio} \& {Barrado y Navascu{\'
  e}s}}{{Mart{\'{\i}}n} et~al.}{2000}]{mart_2000c}
{Mart{\'{\i}}n} E.~L.,  {Brandner} W.,  {Bouvier} J.,  {Luhman} K.~L.,
  {Stauffer} J.,  {Basri} G.,  {Zapatero Osorio} M.~R.,    {Barrado y
  Navascu{\' e}s} D.,  2000, ApJ, 543, 299

\bibitem[\protect\citeauthoryear{{Pinfield}, {Dobbie}, {Jameson}, {Steele},
  {Jones} \& {Katsiyannis}}{{Pinfield} et~al.}{2003}]{pin_2003}
{Pinfield} D.~J.,  {Dobbie} P.~D.,  {Jameson} R.~F.,  {Steele} I.~A.,  {Jones}
  H.~R.~A.,    {Katsiyannis} A.~C.,  2003, MNRAS, 342, 1241

\bibitem[\protect\citeauthoryear{{Pinfield}, {Hodgkin}, {Jameson}, {Cossburn},
  {Hambly} \& {Devereux}}{{Pinfield} et~al.}{2000}]{pin_ple}
{Pinfield} D.~J.,  {Hodgkin} S.~T.,  {Jameson} R.~F.,  {Cossburn} M.~R.,
  {Hambly} N.~C.,    {Devereux} N.,  2000, MNRAS, 313, 347

\bibitem[\protect\citeauthoryear{{Pinfield}, {Hodgkin}, {Jameson}, {Cossburn}
  \& {von Hippel}}{{Pinfield} et~al.}{1997}]{pin_prae}
{Pinfield} D.~J.,  {Hodgkin} S.~T.,  {Jameson} R.~F.,  {Cossburn} M.~R.,
  {von Hippel} T.,  1997, MNRAS, 287, 180

\bibitem[\protect\citeauthoryear{{Pinfield}, {Katsiyannis}, {Mooney},
  {Fitzsimmons} \& {Fletcher}}{{Pinfield} et~al.}{2000}]{pin_t}
{Pinfield} D.~J.,  {Katsiyannis} A.,  {Mooney} C.~J.,  {Fitzsimmons} A.,
  {Fletcher} E.,  2000, IrAJ, 27, 145

\bibitem[\protect\citeauthoryear{{Pinsonneault}, {Stauffer}, {Soderblom},
  {King} \& {Hanson}}{{Pinsonneault} et~al.}{1998}]{pinn_hip}
{Pinsonneault} M.~H.,  {Stauffer} J.,  {Soderblom} D.~R.,  {King} J.~R.,
  {Hanson} R.~B.,  1998, ApJ, 504, 170

\bibitem[\protect\citeauthoryear{{Sanders}}{{Sanders}}{1971}]{sand}
{Sanders} W.~L.,  1971, A\&A, 14, 226

\bibitem[\protect\citeauthoryear{{Tsuji}, {Ohnaka}, {Aoki} \&
  {Nakajima}}{{Tsuji} et~al.}{1996}]{tsuji}
{Tsuji} T.,  {Ohnaka} K.,  {Aoki} W.,    {Nakajima} T.,  1996, A\&A, 308, L29

\bibitem[\protect\citeauthoryear{{Wang}, {Chen}, {Zhao} \& {Jiang}}{{Wang}
  et~al.}{1995}]{wang}
{Wang} J.~J.,  {Chen} L.,  {Zhao} J.~H.,    {Jiang} P.~F.,  1995, A\&AS, 113,
  419

\end{thebibliography}
\appendix
\section{Tables of photometry, membership and binarity criteria, and mass estimates}
\begin{table*}
\caption{Photometry of the WFC candidates that were initially considered interesting (see Section 3). Final photometric membership criteria are given in the last 3 columns. $\ddag$ and $\dagger$ are also \textit{Riz}1 and 2MASSJ08350622+1953050 respectively.} \label{tab:photom} 
\
\begin{tabular}{cccccccccc}
\hline
&&&&&&&\multicolumn{2}{c}{Membership criteria} \\
\multicolumn{1}{c}{WFC} & \multicolumn{1}{c}{RA (J2000.0)}& \multicolumn{1}{c}{Dec. (J2000.0)}& \multicolumn{1}{c}{I$_c$} & \multicolumn{1}{c}{J} & \multicolumn{1}{c}{H}
& \multicolumn{1}{c}{K} & \multicolumn{1}{c}{IK} & \multicolumn{1}{c}{JK} 
& \multicolumn{1}{c}{JHK}  \\
\hline
11 & 8   45   41.05  & 19  38   02.6 & 21.37 &18.37 $\pm 0.04$ & 17.51 $\pm 0.02 $ & 16.84  $\pm 0.02 $ & Y & Y & Y \\
22 & 8 44 34.29 &  19 14 23.3 & 21.20 &18.37 $\pm 0.04$ & 17.74 $\pm 0.06 $ & 17.22  $\pm 0.04 $ & N? & N? & ? \\
24 &  8 38 32.13  & 19 59 27.1   & 21.19 &17.99 $\pm 0.03$ & 17.30 $\pm 0.02 $ & 16.78  $\pm 0.02 $ & Y & Y & Y \\
30  &   8 41 56.61  & 19 20 34.8 & 21.13 &17.91 $\pm 0.04$ & 17.45 $\pm 0.03 $ & 17.03  $\pm 0.04 $ & N? & N & ? \\
53 &   8 46 39.19  & 19 27 37.0  & 20.90 &17.83 $\pm 0.03$ & 17.18 $\pm 0.03 $ & 16.64  $\pm 0.03 $ & Y & Y & Y \\
57 &   8 37 19.80 &  19 38 51.4  &20.79 &18.82 $\pm 0.06$ & 17.90 $\pm 0.06 $ & 16.91  $\pm 0.03 $ & N? & N & N \\
60 &  8 41 25.03  & 19 09 25.0 &  20.79 &17.61 $\pm 0.02$ & 16.80 $\pm 0.02 $ & 16.22  $\pm 0.01 $ & Y & Y & Y \\
76 &  8 37 27.61 &  19 37 33.0 & 20.51 &17.58 $\pm 0.01$ & 16.83 $\pm 0.01 $ & 16.46  $\pm 0.02 $ & Y & Y & ? \\
81 &   8 41 50.21  & 19 06 18.4  &  20.33 &17.47 $\pm 0.01$ & 16.80 $\pm 0.01 $ & 16.30  $\pm 0.01 $ & Y & Y & Y \\
87 &  8 38 02.34  & 19 46  09.0 & 19.83 &17.14 $\pm 0.01$ & 16.64 $\pm 0.01 $ & 16.06  $\pm 0.01 $ & Y & Y & ? \\
88  &   8 40 43.86  & 19 06 00.8 & 19.83 & - & - &   15.59$\pm0.02 $ & Y & - & - \\
91$^\ddag$ &   8 36 54.46  & 19 54 15.3  &19.75 &17.07 $\pm 0.01$ & 16.47 $\pm 0.01 $ & 15.99  $\pm 0.01 $ & Y & Y & Y \\
94 &   8 35 51.93  & 19 20 33.6 & 19.69 &17.04 $\pm 0.01$ & 16.42 $\pm 0.01 $ & 15.98  $\pm 0.01 $ & Y & Y & Y \\
103$^\dagger$ &  8 35 06.12  & 19 53 05.3  &  19.56 & - & - &   14.28$\pm0.01 $ & N & - & - \\
104 &   8 41 47.75  & 19 48 37.7  &  19.56 & - & - &   15.80$\pm0.02 $ & Y & - & - \\
115 &   8 36 44.74  & 19 17 23.9  & 19.38 & - & - &   15.93$\pm0.03 $ & Y & - & - \\
118 &    8 37 38.73 &  19 03 09.6  & 19.30 & - & - &   15.53$\pm0.02 $ & Y & - & - \\
135 &   8 38  07.96  & 19 43 28.6 & 19.11 & - & - &   15.74$\pm0.02 $ & Y & - & - \\
136 &    8 45 39.10  & 19 38 04.2 & 19.09 & - & - &   15.80$\pm0.02 $ & Y & - & - \\
138 &    8 36 52.03  & 19 00 42.2 & 19.08 & - & - &   15.78$\pm0.08 $ & Y & - & - \\
140 &   8 36 21.20  & 19 38 45.5  & 19.06 & - & - &   15.56$\pm0.02 $ & Y & - & - \\
142 &   8 34 54.28  & 19 02 32.7 & 19.03 & - & - &   15.62$\pm0.02 $ & Y & - & - \\
150 &    8 36 48.06  & 19 49 02.2  & 18.92 & - & - &   15.52$\pm0.02 $ & Y & - & - \\
159 &   8 37 41.90 &  19 09 31.0  & 18.82 & - & - &   15.59$\pm0.02 $ & Y & - & - \\
163 &   8 40 53.59  & 19 40 58.3  & 18.78 & - & - &   15.60$\pm0.02 $ & Y & - & - \\
169 &   8 36 55.63  & 19 36 15.5 & 18.69 & - & - &   14.99$\pm0.01 $ & Y & - & - \\
191 &   8 38 11.96  & 19 59 43.5 & 18.25 & - & - &   15.22$\pm0.02 $ & Y & - & - \\
193 &   8 41 27.76  & 19 07 19.7  & 18.21 & - & - &   15.27$\pm0.02 $ & Y & - & - \\
200 &     8 46 16.82 &  19 52 30.2    &  18.17 & - & - &   15.05$\pm0.02 $ & Y & - & - \\
204 &   8 36 29.90  & 19 38 37.5  & 18.08 & - & - &   15.14$\pm0.02 $ & Y & - & - \\
207 &   8 39 39.37  & 19 47 54.9 & 18.04 & - & - &   15.19$\pm0.02 $ & Y & - & - \\
208 &   8 40 50.31  & 18 23 59.8  & 18.02 & - & - &   14.82$\pm0.01 $ & Y & - & - \\
210 &   8 38 50.24  & 19 30 18.6  & 18.02 & - & - &   15.14$\pm0.02 $ & Y & - & - \\
216 &  8 46 48.87 &  19 32  09.4  &  17.93 & - & - &   14.98$\pm0.02 $ & Y & - & - \\
220 &   8 37 02.23  & 19 52 07.6  & 17.88 & - & - &   14.76$\pm0.01 $ & Y & - & - \\
\hline
\end{tabular} 
\end{table*}
\pagebreak
\newpage
\bigskip 
\begin{table*}
\caption{HSHJ photometric non-members. HSHJ\{2,8,9,11,39,54,73,88,109,110,217,413\} had no 2MASS counterpart ($\leq 3 \arcsec$). HSHJ\{59,114,294\} were flagged in 2MASS as contaminated, due most likely to either photometric confusion from nearby sources or to diffraction spikes. HSHJ\{509,510,515\} have red colours typical for foreground field dwarfs but, in light of the fact that they all appear on the same HSHJ plate, we consider their red colours as potentially suspect, and do not flag them as non-members.}
\label{tab:photom_hshj} 
\begin{tabular}{ccccc}
\hline
\multicolumn{1}{c}{HSHJ} & \multicolumn{1}{c}{RA (J2000.0)}& \multicolumn{1}{c}{Dec. (J2000.0)}& \multicolumn{1}{c}{I$_c$} & \multicolumn{1}{c}{K} \\
\hline
 
 46  & 8 33 43.55 & 19 56 34.94 & 17.26  & 14.94 $\pm 0.10$ \\
 98 & 8 36 05.97 & 22 26 15.98 & 12.62  & 11.89 $\pm 0.02$ \\
174 & 8 37 35.96 & 18 41 25.25 & 16.78 & 14.55 $\pm 0.07$ \\
199 &  8 38 16.04 & 21 23 29.87 & 12.63 & 11.44 $\pm 0.02$ \\
200 &  8 38 14.54 & 19 52 34.41 & 15.17 & 13.17 $\pm 0.03$ \\
208 & 8 38 36.26 & 22 32 27.56 & 12.62  & 11.64 $\pm 0.02$ \\
231 & 8 38 52.07 & 18 07 44.53 & 12.89  & 11.61 $\pm 0.02$ \\
315 & 8 40 21.35 & 19 10 53.35 & 12.20  & 10.75 $\pm 0.03$ \\
323 & 8 40 33.53 & 19 38 00.48 & 11.16  & 10.17 $\pm 0.02$ \\
332 & 8 40 53.87 & 19 59 55.45 & 13.99  & 12.10 $\pm 0.02$ \\
336 & 8 40 55.69 & 19 51 51.84 & 11.88  & 10.79 $\pm 0.02$ \\
387 & 8 41 44.13 & 22 04 17.41 & 16.11  & 13.98 $\pm 0.06$ \\
391 & 8 41 47.76 & 19 24 43.72 & 11.00  & 10.09 $\pm 0.02$ \\
437 & 8 43 04.29 & 19 17 33.76 & 11.62  & 10.56 $\pm 0.02$ \\
469 & 8 43 51.30 & 19 23 54.15 & 17.12  & 14.80 $\pm 0.12$ \\
477 & 8 44 17.16 & 19 51 55.64 & 12.77  & 11.12 $\pm 0.02$ \\
480 & 8 44 18.25 & 18 43 50.74 & 12.74  & 11.48 $\pm 0.02$ \\
487 & 8 44 29.75 & 19 28 03.43 & 11.40  & 10.59 $\pm 0.02$ \\

\hline
\end{tabular} 
\end{table*}
\pagebreak

\begin{table*}
\caption{Praesepe photometric candidates found in our \textit{cluster sequence region} (see text) in ascending \textit{I}-\textit{K}. The Table includes WFC candidates (from this work), \textit{Riz} and \textit{Iz} candidates from P03, and RPr1 \citep{mag}. Mass estimates (0.001 M$_\odot$) are given using both 0.5 and 1.0 Gyr relations, with NEXTGEN and Dusty models providing the lower and upper limits respectively. Where possible, mass estimates were made using the more reliable \textit{J-} band predictions, otherwise the \textit{K-} band was used. Unresolved binaries are flagged (see \S \ref{subsec:unres_bin}). }                                                                                                                  
\label{tab:photom_red} 
\begin{tabular}{|c|c|c|c|c|c|c|c|c|c|}
\hline

\multicolumn{1}{|c|}{Candidate} & \multicolumn{1}{c}{\textit{I$_c$}}& \multicolumn{1}{|c|}{\textit{K}}& \multicolumn{1}{|c|}{\textit{I$_c$}-\textit{K}} & \multicolumn{1}{|c|}{\textit{J}-\textit{H}} & \multicolumn{1}{|c|}{\textit{H}-\textit{K}} & \multicolumn{1}{|c|}{Mass (0.5 Gyr)} & \multicolumn{1}{c|}{Mass (1.0 Gyr)} & \multicolumn{1}{|c|}{Bin}\\
\hline
 
WFC138 &   19.08 &    15.78  &   3.30  &  -- &  -- & 93--88 & 97--92\\
WFC135 &   19.11 &    15.74  &   3.37  &  -- &    -- & 95--89 & 97--93\\
WFC150 &   18.92 &    15.52  &   3.40  &  -- &   -- & 101--96 & 103--98\\
WFC142 &   19.03 &    15.62  &   3.42  &  -- &  -- & 99--92 & 101--96\\
RIZ2 &   18.19 &    14.77  &   3.42  &  0.56 &  0.34 & & & Y\\
WFC115 &   19.38 &    15.93  &   3.44  &  -- &   -- & 89--83 & 93--89\\
IZ93 &   19.20 &    15.75  &   3.45  &  0.55 &   0.46 & 94--88 & 97--92\\
IZ126 &   18.70 &    15.22  &   3.48  &  0.37 &  0.35 & & &Y\\
WFC140 &   19.06 &    15.56  &   3.50  &  -- &    -- & 100--94 & 102--97\\
RIZ23 &   19.06 &    15.53  &   3.53  &  0.59 &   0.51 &  & &Y\\
IZ23 &   19.10 &    15.57  &   3.53  &  0.43 &   0.47 & 94--93 & 97--96\\
RIZ11 &   19.47 &    15.84  &   3.63  &  0.43 &   0.39 & 88--87 & 92--91\\
IZ36 &   19.31 &    15.66  &   3.65  &  0.51 &   0.45 & 90--89 & 93--92\\
WFC169 &   18.69 &    14.99  &   3.70  &  -- &  -- &  & &Y \\
RIZ21 &   18.73 &    15.03  &   3.70  &  0.50 &  0.36 & & &Y \\
WFC94 &   19.69 &    15.98  &   3.71  &  0.61 &   0.44 & 79--78 & 86--85\\
WFC91 &   19.75 &    15.99  &   3.75  &  0.60 &   0.48 & 78--77 & 86--85\\
WFC104 &   19.56 &    15.80  &   3.76  &  -- &   -- & 92--87 & 96--92\\
WFC118 &   19.30 &    15.53  &   3.77  &  -- &   -- & & &Y \\
WFC87 &   19.83 &    16.06  &   3.77  &  0.50 &   0.58 & 76--75 & 85--84\\ 
WFC81 &   20.33 &    16.30  &   4.03  &  0.67 &   0.49 & 70--69 & 80--79\\
WFC76 &   20.51 &    16.46  &   4.05  &  0.75 &   0.37 & 68--66 & 79--78\\
RIZ18 &   19.63 &    15.40  &   4.23  &  0.66 &  0.34 & & &Y \\
 WFC88 &   19.83 &  15.59  &   4.24  &  -- &  -- & & &Y \\
RIZ24 &   20.43 &    16.17  &   4.26  &  0.62 &   0.51 & 82--77 & 88--86\\
WFC53 &   20.90 &    16.64  &   4.26  &  0.65 &   0.54 & 64--63 & 78--76\\
WFC24 &   21.19 &    16.78  &   4.41  &  0.68 &   0.53 & 60--63  & 73--76 &\\
WFC11 &   21.37 &    16.83  &   4.54  &  0.85 &   0.68 & 57--56 & 74--71\\
WFC60 &   20.79 &    16.22  &   4.56  &  0.80 &   0.58 & & &Y \\
RPr1 & 21.01 & 16.42 & 4.60 & -- & -- & 70--67   & 80--78  \\

\hline
\end{tabular} 
\end{table*}

\label{lastpage}

\end{document}